\documentclass[journal, 10pt]{IEEEtran}
\normalsize
\usepackage{graphicx,epsfig,epstopdf,amsmath,amssymb,enumerate,float,color,cite}
\usepackage{subfig,stfloats,multicol}
\usepackage[labelsep=period,font=small]{caption}

\UseRawInputEncoding

\ifCLASSINFOpdf
\else
\fi
\begin{document}
\setlength{\textfloatsep}{4.5pt}
\title{A General 3D Non-Stationary Wireless Channel Model for 5G and Beyond}

\author{Ji Bian,~\IEEEmembership{Member,~IEEE}, Cheng-Xiang Wang,~\IEEEmembership{Fellow,~IEEE}, Xiqi Gao,~\IEEEmembership{Fellow,~IEEE}, Xiaohu~You,~\IEEEmembership{Fellow,~IEEE}, and Minggao Zhang

\thanks{Manuscript received January 31, 2020; revised August 11, 2020 and
December 12, 2020; accepted December 23, 2020.
The authors would like to acknowledge the support from the National Key R\&D Program of China under Grant 2018YFB1801101, the National Natural Science Foundation of China (NSFC) under Grant 61960206006,
the Shandong Provincial Natural Science Foundation for Young Scholars
of China under Grant ZR2020QF001, Taishan Scholar Program of Shandong Province, the Frontiers Science Center for Mobile Information Communication and Security, the High Level Innovation and Entrepreneurial Research Team Program in Jiangsu, the High Level Innovation and Entrepreneurial Talent Introduction Program in Jiangsu, the Research Fund of National Mobile Communications Research Laboratory, Southeast University, under Grant 2020B01, the Fundamental Research Funds for the Central Universities under Grant 2242020R30001, and the EU H2020 RISE TESTBED2 project under Grant 872172.}

\thanks{J. Bian is with School of Information Science and Engineering, Shandong Normal University, Jinan, Shandong, 250358, China (e-mail: jibian@sdnu.edu.cn).
}
\thanks{ C.-X. Wang (corresponding author), X. Q. Gao and X. H. You are with the National Mobile Communications Research Laboratory, School of Information Science and Engineering, Southeast University, Nanjing, 210096, China, and also with the Purple Mountain Laboratories, Nanjing, 211111, China (email: \{chxwang, xqgao, xhyu\}@seu.edu.cn).
}
\thanks{M. G. Zhang is with Shandong Provincial Key Lab of Wireless Communication Technologies, School of Information Science and Engineering, Shandong University, Qingdao, Shandong, 266237, China (e-mail: zmg22s@163.com).
}
}

\markboth{IEEE Transactions on WIRELESS COMMUNICATIONS,~Vol.~XX, No.~XX, MONTH~2020}%
{Submitted paper}

\maketitle
\begin{abstract}
In this paper, a novel three-dimensional (3D) non-stationary geometry-based stochastic model (GBSM) for the fifth generation (5G) and beyond 5G (B5G) systems is proposed.
The proposed B5G channel model (B5GCM) is designed to capture various channel characteristics in (B)5G systems such as space-time-frequency  (STF) non-stationarity, spherical wavefront (SWF), high delay resolution, time-variant velocities and directions of motion of the transmitter, receiver, and scatterers, spatial consistency, etc.
By combining different channel properties into a general channel model framework, the proposed B5GCM is able to be applied to multiple frequency bands and multiple scenarios, including massive multiple-input multiple-output (MIMO), vehicle-to-vehicle (V2V), high-speed train (HST), and millimeter wave-terahertz (mmWave-THz) communication scenarios.
Key statistics of the proposed B5GCM are obtained and compared with those of standard 5G channel models and corresponding measurement data, showing the generalization and usefulness of the proposed model.
\end{abstract}


\IEEEpeerreviewmaketitle

\section{Introduction}
The growing requirement of high data rate transmission caused by the popularization of wireless services and applications results in a spectrum crisis in current sub-6~GHz bands.
To address this challenge, the fifth generation (5G)/beyond 5G (B5G) wireless communication systems will transmit data using millimeter wave (mmWave)/terahertz (THz) bands in multiple propagation scenarios, e.g., high-speed train (HST) and vehicle-to-vehicle (V2V) scenarios \cite{Akyildiz2014Terahertz}.
The short wavelengths of mmWave-THz bands make it possible to deploy large antenna arrays with high beamforming gains that can overcome the severe path loss \cite{You2017BDMA}.
Revolutionary technologies employed in 5G and B5G wireless communication systems such as massive multiple-input multiple-output (MIMO), HST, V2V, and mmWave-THz communications introduce new channel properties, such as spherical wavefront (SWF), spatial non-stationarity, oxygen absorption, etc.
These in turn will set new requirements to standard (B)5G channel models, i.e., supporting multiple frequency bands and multiple scenarios, as follows \cite{METISD1.4,Wang2018Survey}:
\begin{enumerate}
\item Multiple frequency bands, covering sub-6~GHz, mmWave, and THz bands;
\item Large bandwidth, e.g., 0.5--4~GHz for mmWave bands and 10~GHz for THz bands;
\item Massive MIMO scenarios: spatial non-stationarity and SWF;
\item HST scenarios: temporal non-stationarity including parameters' drifting and clusters' appearance and disappearance over time;
\item V2V scenarios: temporal non-stationarity and multi-mobility, i.e., the transmitter (Tx), receiver (Rx), and scatterers may move with time-variant velocities and heading directions;
\item Three-dimensional (3D) scenarios, especially for indoor and outdoor small cell scenarios;
\item Spatial consistency scenarios, i.e., closely located links experience similar channel statistical properties.
\end{enumerate}

In massive MIMO communications, the large arrays make the channel spatially non-stationary, which means channel parameters and statistical properties vary along array axis \cite{Wang2018Survey}.
For example, measurement results in \cite{Payami2012MaMIMO_meas} and \cite{Li2019MaMIMO} showed that the angles of multipath components (MPCs) drift across the array, justifying the SWF assumption of the channel.
This implies that travel distances from every Tx antenna element to the Rx/scatterers at each time instant (if temporal non-stationarity is considered) have to be calculated, resulting in high model complexity \cite{Wu2015,Wu2014,Wu2017unified}.
Measurements in \cite{Gao2015MaMIMO} and \cite{Ai2017MaMIMO} revealed that the mean delay, delay spread, and cluster power can vary across the large array.
Here, a cluster is a group of MPCs having similar properties in delay, power, and angles.
Note that the cluster power variation over array has not been considered in most massive MIMO channel models \cite{Wu2015,Wu2014,Wu2017unified}.
Furthermore, when large arrays are adopted, clusters illustrate a partially visible property.
Some clusters are visible over the whole array, while other clusters can only interact with part of the array \cite{Gao2013MaMIMO_meas,Li2019MaMIMO}.
In \cite{Li2019MaMIMO_TWC} and \cite{Flordelis2020MaMIMO}, the partially visibility of clusters was modeled by introducing the concept of ``BS-visibility region (VR)".
Other researches such as \cite{Wu2015} and \cite{Carlos2017MaMIMO} described the partial visibility of clusters using birth-death or Markov processes.
In general, how to efficiently and synthetically model the non-stationarities in the time and space domains has to be solved in the (B)5G channel modeling.

In V2V channels, channel parameters and statistical properties are time-varying caused by the motions of the Tx, Rx, and scatterers \cite{He2015HSTnonWSS}.
Besides, channel measurements showed that clusters of V2V channels can exhibit a birth-death behavior, i.e., appear, exist for a time period, and then disappear \cite{Wang2018Survey}.
Large numbers of V2V channel models were designed as pure geometry-based stochastic model (pure-GBSM) \cite{Yuan2015V2V}, \cite{Zhao2016V2V}.
The scatterers of those models were assumed to be located on regular shapes, which are less versatile than those of the WINNER/3GPP models \cite{WINNERPlus,3GPP38901}.
More realistic V2V channel models, e.g., \cite{Yang2019V2V} and \cite{Yang2020V2V}, were developed based on channel measurements.
However, the motions of scatterers were neglected.
Furthermore, the variations of velocity and trajectory of the Tx/Rx
were not taken into account.

The HST channels share some similar properties with V2V channels, e.g., large Doppler shifts and temporal non-stationarity.
Widely used standard channel models, e.g., WINNER~II \cite{WINNERII} and IMT-Advanced channel models \cite{IMTA} can be applied to HST scenarios where the velocity of train can be up to 350~km/h.
However, those models are developed on temporal wide-sense stationary (WSS) assumption.
The HST channel model in \cite{Ghazal2017IMT-A} was developed based on IMT-Advanced channel model \cite{IMTA} by taking into account the time-varying angles and cluster evolution.
More general HST channel model was proposed in \cite{Bi2019HST}, which extended the elliptical model by assuming velocity and moving direction variations.
However, the above-mentioned models are two-dimensional (2D) and can only be applied to the scenarios where the transceiver and scatterers are sufficiently far away.
In \cite{Liu2020Tunnel}, a 3D non-stationary HST channel model was proposed, which can only be used in tunnel scenarios.
In \cite{Yang2019HST}, a tapped delay line model was presented for various HST scenarios, e.g., viaduct and cutting.
The fidelity of the model relies on the model parameters obtained from ray-tracing, resulting in high computation complexity.

For the mmWave-THz communications, high frequencies result in large path loss and render the mmWave-THz propagation susceptible to blockage effects and oxygen absorption \cite{METISD1.4}.
Compared with sub-6~GHz bands, the mmWave-THz channels are sparser \cite{He2020mmWave}.
High delay resolution is required in channel modeling due to the large bandwidth.
Rays within a cluster can have different time of arrival, leading to unequal ray powers \cite{3GPP38901}.
Besides, as the relative bandwidth increases, the channel becomes non-stationary in the frequency domain, which means the uncorrelated scattering (US) assumption may not be fulfilled \cite{Molisch2005UWB}.
Furthermore, directional antennas are often deployed to overcome the high attenuation at mmWave-THz frequency bands.
In \cite{He2018M2M}, a 2D mmWave V2V channel model was proposed.
The influence of directional antennas was represented by eliminating the clusters outside the main lobe of antenna patterns.
MmWave-THz channel models should faithfully recreate the spatial, temporal and frequency characteristics for every single ray, such as the models in \cite{Samimi2016mmW} and \cite{He2017THz}.
However, those models are oversimplified and cannot support time evolution since the model parameters are time-invariant.

Apart from modeling channel characteristics in various scenarios, another challenge for (B)5G channel modeling lies in how to combine those channel characteristics into a general modeling framework.
Standard channel models aim to solve this problem.
In order to achieve smooth time evolution, the COST 2100 channel model introduces the ``VR", which indicates whether a cluster can be ``seen" from the mobile station (MS).
The clusters are considered to physically exist in the environments and do not belong to a specific link.
Thus, closely located links can experience similar environments, justifying the spatial consistency of the model.
However, the COST 2100 channel model can only support sub-6~GHz bands.
Furthermore, massive MIMO and dual-mobility were not supported.
The QuaDriGa \cite{QuaDRiGa_Doc} channel model is extended from 3GPP TR36.873 \cite{3GPP3D} and support frequencies over 0.45--100~GHz.
The model parameters were generated based on spatially correlated random variables.
Links located nearby share similar channel parameters and therefore supports spatial consistency.
However, the complexities of those models are relatively high and the dual-mobility was neglected.
The 3GPP TR38.901 \cite{3GPP38901} channel model extended the 3GPP TR36.873 by supporting the frequencies over 0.5--100~GHz.
The oxygen absorption and blockage effect for the mmWave bands were modeled.
However, for high-mobility scenarios, the clusters fade in and fade out were not considered, which results in limited ability for capturing time non-stationarity.
Besides, the dual-mobility and SWF cannot be supported.
Note that all the above-mentioned standard channel models assumed constant cluster power for different antenna elements and did not consider scatterers movements.

A GBSM called more general 5G channel model (MG5GCM) aims at capturing various channel properties in 5G systems \cite{Wu2017unified}.
Based on a general model framework, the model can support various communication scenarios.
However, the azimuth angles, elevation angles, and travel distances between Tx/Rx and scatterers were generated independently.
The model can only evolve along the time and array axes, and neglected the non-stationary properties in the frequency domain.
The locations of scatterers were implicitly determined by the angles and delays, which makes the model difficult to achieve spatial consistency.
Besides, it neglected the dynamic velocity and direction of motion of the Tx, Rx, and scatterers.

Through above analysis, we find that none of these standard channel models can meet all the 5G channel modeling requirements.
Considering these research gaps, this paper proposes a general (B)5G channel model (B5GCM) towards multiple frequency bands and multiple scenarios.
The \textbf{contributions} of this paper are listed as follows:
\begin{enumerate}
\item
    The system functions, correlation functions (CFs), and power spectrum densities (PSDs) of space-time-frequency (STF) stationary and non-stationary channels are presented.
    A general 3D STF non-stationary ultra-wideband channel model is developed.
    By setting appropriate channel model parameters, the model can support multiple frequency bands and multiple scenarios.
\item
    A highly accurate approximate expression of the 3D SWF is proposed, which is more scalable and efficient than the traditional modeling method and can capture spatial and temporal channel non-stationarities.
    The cluster evolutions along the time and array axes are jointly considered and simulated using a unified birth-death process.
    The smooth power variation over the large arrays is modeled by a 2D spatial lognormal process.
\item
    The novel ellipsoid Gaussian scattering distribution is proposed which can jointly describe the azimuth angles, elevation angles, and distances from the Tx(Rx) to the first(last)-bounce scatterers.
    By tracking the locations of the Tx, Rx, and scatterers, the spatial consistency of the channel is supported.
\item
    The proposed B5GCM takes into account a multi-mobility communication environment, where the Tx, Rx, and scatterers can change their velocities and moving directions.
\item
    Key statistics including local STF-CF, local spatial-Doppler PSD,  local Doppler spread, and array coherence distance are derived and compared with standard 5G channel models and the corresponding measurement data.
\end{enumerate}

The remainder of this paper is organized as follows.
In Section II, the system, correlation, and spectrum functions of the STF non-stationary channel model are presented.
The proposed B5GCM is described in Section III.
Statistics of the B5GCM are investigated in Section IV.
In Section V, numerical and simulation results are provided and
discussed.
Conclusions are drawn in Section VI.

\section{System Functions, CFs, and PSDs of Space, Time, and Frequency Non-Stationary Channels}
\begin{figure}
\centering\includegraphics[width=3.3in]{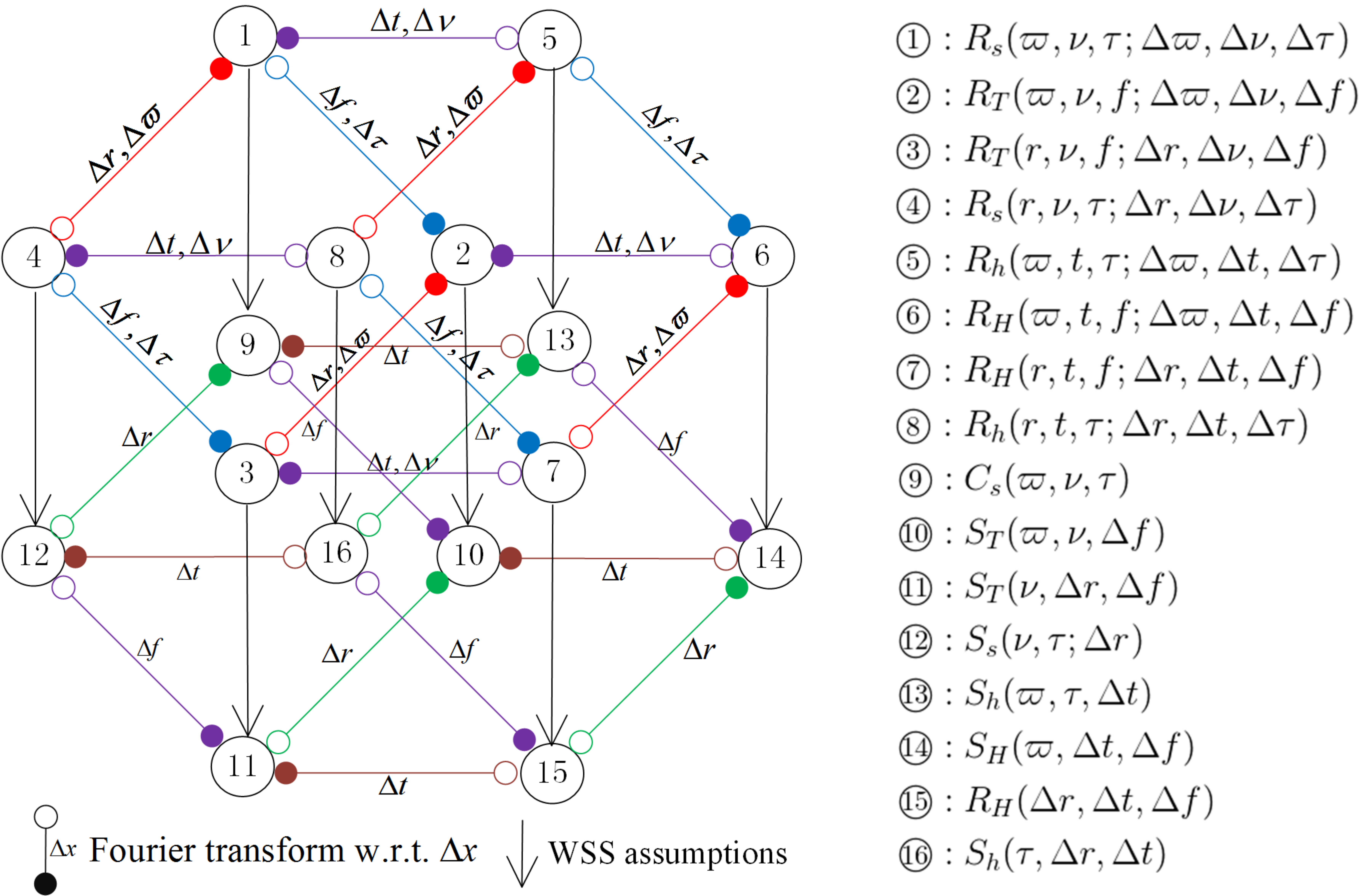}
\caption{The relationship among CFs for STF-WSS and STF-non-WSS channels.}
\label{fig_CorrelationFunctions}
\end{figure}

Wireless channels can be described through system functions, CFs, and PSDs.
The time-frequency selectivity and delay-Doppler dispersion of wireless channels have been studied in \cite{Bello1963} and \cite{Matz2005nonWSSUS}.
In this paper, the non-stationarity of channels in the space domain is considered, which is an essential property for massive MIMO channels.
The basic function characterizing the wireless channel is the space and time-varying channel impulse response (CIR) $h(r, t,\tau)$.
It is modeled as a stochastic process on space $r$, time $t$, and delay $\tau$.
Note that the space domain indicates the region where $r$ is confined along a linear antenna array.
Taking the Fourier transform of $h(r,t,\tau)$ with respect to (w.r.t.) $\tau$ results in space- and time-varying transfer function, which is given by
\begin{equation}
H(r,t,f)=\int h(r,t,\tau)e^{-j2\pi\tau f}{\rm d}\tau.
\end{equation}
The space and time-varying transfer function characterizes the space, time, and frequency selectivity of wireless channels.
Moreover, taking the Fourier transform of $h(r,t,\tau)$ w.r.t. $r$ and $t$ results in the spatial-Doppler Doppler delay spread function, i.e.,
\begin{equation}\label{equ_SpreadFun}
s(\varpi,\nu,\tau)=\iint h(r,t,\tau)e^{-j2\pi(r\varpi\lambda^{-1}+t\nu)}{\rm d}r {\rm d}t.
\end{equation}
Here, $\lambda$ is the wavelength, $\varpi$ is the spatial-Doppler frequency variable defined as $\varpi=\Omega \cdot \tilde \Omega$, where $\Omega$ is an angle unit vector of the departure/arrival waves, $\tilde \Omega$ indicates the antenna array orientation \cite{Fleury2000SpatialDoppler}.
The space variable $r$ and the spatial-Doppler variable $\varpi$ are Fourier transformation pair.
Since $r$ has a distance unit, the unit of $\varpi$ must be its reciprocal, i.e., per normalized distance (w.r.t. $\lambda$).
The terminology $\varpi$ stems from the fact that $\Omega \cdot \tilde \Omega$ is the Doppler shift of a wave with direction $\Omega$ impinging on an antenna which moves with $\lambda$ m/s in direction $\tilde \Omega$.
The spatial-Doppler Doppler delay spread function characterizes the dispersions of wireless channels in the spatial-Doppler frequency, Doppler frequency, and delay domains.
Similarly, by taking the Fourier transform of $h(r,t,\tau)$ w.r.t. $r$, $t$, and/or $\tau$, totally eight system functions can be obtained \cite{Kattenbach2002SpatialDoppler}.
Based on those formulas, e.g., $h(r,t,\tau)$, $H(r,t,f)$, and $s(\varpi,\nu,\tau)$, the six-dimensional (6D) CFs are derived as
\begin{align}
\label{equ_R_h}
&R_h(r,t,\tau; \Delta r,\Delta t,\Delta \tau)\nonumber \\  &=\mathbb E\{h(r,t,\tau)h^*(r-\Delta r,t-\Delta t,\tau-\Delta \tau)\} \\
\label{equ_STF_CorFun}
&R_H(r,t,f;\Delta r,\Delta t,\Delta f)\nonumber \\  & =\mathbb E\{H(r,t,f)H^*(r-\Delta r,t-\Delta t,f-\Delta f)\} \\
\label{equ_R_s}
&R_s(\varpi,\nu,\tau;\Delta \varpi, \Delta \nu,\Delta \tau)\nonumber \\  &=\mathbb E\{s(\varpi,\nu,\tau)s^*(\varpi-\Delta \varpi,\nu-\Delta \nu,\tau-\Delta \tau)\}
\end{align}
where $\mathbb E\{\cdot\}$ indicates ensemble average and $(\cdot)^*$ stands for complex conjugation, $\Delta r$, $\Delta t$, and $\Delta f$ are space, time, and frequency lags, respectively.
A channel is STF non-stationary if $R_H(r,t,f;\Delta r,\Delta t,\Delta f)$ is not only a function of $\Delta r$, $\Delta t$, and $\Delta f$, but also relies on $r$, $t$, and $f$.
Its simplification, i.e., WSS over $r$, $t$, and $f$ is widely used in the existing channel models.
However, it is valid only if the channel satisfies certain conditions.
For example, when the distance from the Tx to the Rx (or a cluster) is less than the Rayleigh distance, i.e.,  $\frac{2L^2}{\lambda}$, where $L$ denotes the aperture size of the antenna array and $\lambda$ is the carrier wavelength, the spatial WSS condition is fulfilled \cite{Wu2014}.
The temporal WSS assumption is valid as long as the channel stationary interval is larger than the observation time.
Finally, when the relative bandwidth of the channel is small (typically less than 20\% of the carrier frequency), the channel becomes WSS in the frequency domain.
Considering the STF WSS conditions, the 6D CFs in \eqref{equ_R_h}-\eqref{equ_R_s} can be reduced to
\begin{align}
\label{equ_CorFun1}
&R_h(r,t,\tau; \Delta r,\Delta t,\Delta \tau)=S_h(\tau;\Delta r, \Delta t)\delta(\Delta \tau) \\
\label{equ_CorFun2}
&R_H(r,t,f;\Delta r,\Delta t,\Delta f)=R_H(\Delta r,\Delta t,\Delta f) \\
\label{equ_CorFun3}
&R_s(\varpi,\nu,\tau;\Delta \varpi,\Delta \nu,\Delta \tau)\nonumber \\&= C_s(\varpi,\nu,\tau)\delta(\Delta \varpi)\delta(\Delta \nu)\delta(\Delta \tau)
\end{align}
where $\delta (\cdot)$ is the Dirac function, $S_h(\tau;\Delta r, \Delta t)$ is space time correlation, delay PSD and $C_s(\varpi,\nu,\tau)$ is spatial-Doppler Doppler delay PSD.
Fig. \ref{fig_CorrelationFunctions} shows the complete CFs and their simplifications by the STF WSS assumption.

Another important function is the STF-varying spatial-Doppler Doppler delay PSD, since it plays a central role in deriving other correlation/spectrum functions.
For example, (\ref{equ_R_h})--(\ref{equ_R_s}) are written as
\begin{align}
&R_h(r,t,\tau; \Delta r,\Delta t,\Delta \tau)\nonumber \\ &=\iiint C_s(r,t,f;\varpi,\nu,\tau)e^{j2\pi(\varpi \Delta r\lambda^{-1}+\nu\Delta t +f\Delta\tau )}  {\rm d} \varpi {\rm d}\nu {\rm d}f
\\
&R_H(r,t,f;\Delta r,\Delta t,\Delta f)\nonumber \\ &=\iiint C_s(r,t,f;\varpi,\nu,\tau) e^{j2\pi(\varpi\Delta r\lambda^{-1}+\nu\Delta t-\tau\Delta f)} {\rm d}\varpi  {\rm d}\nu {\rm d} \tau
\\
&R_s(\varpi,\nu,\tau;\Delta \varpi,\Delta \nu,\Delta \tau)\nonumber \\ &=\iiint C_s(r,t,f;\varpi,\nu,\tau) e^{j2\pi (-\Delta \varpi r\lambda^{-1}-\Delta \nu t+\Delta \tau f)} {\rm d} r {\rm d} t {\rm d} f.
\end{align}
The STF-varying spatial-Doppler PSD, Doppler PSD, and delay PSD can be obtained by integrating $C_s(r,t,f;\varpi,\nu,\tau)$ over other two dispersion domains, and are expressed as
\begin{align}
G_s(r,t,f;\varpi )=\iint C_s(r,t,f;\varpi,\nu,\tau){\rm d}\nu {\rm d}\tau
\\
Q_s(r,t,f;\nu)=\iint C_s(r,t,f;\varpi,\nu,\tau){\rm d}\varpi {\rm d}\tau
\\
P_s(r,t,f;\tau)=\iint C_s(r,t,f;\varpi,\nu,\tau){\rm d}\varpi {\rm d}\nu.
\end{align}
The STF-varying spatial-Doppler PSD, Doppler PSD, and delay PSD describe the average power distribution at space $r$, time $t$, and frequency $f$ over the spatial-Doppler, Doppler, and delay domains, respectively.
Note that STF-varying delay PSD $P_s(r,t,f;\tau)$ is also called STF-varying power delay profile (PDP).

\section{The 3D Non-Stationary Ultra-Wideband Massive MIMO GBSM}
\begin{figure}
\centering\includegraphics[width=3.5in]{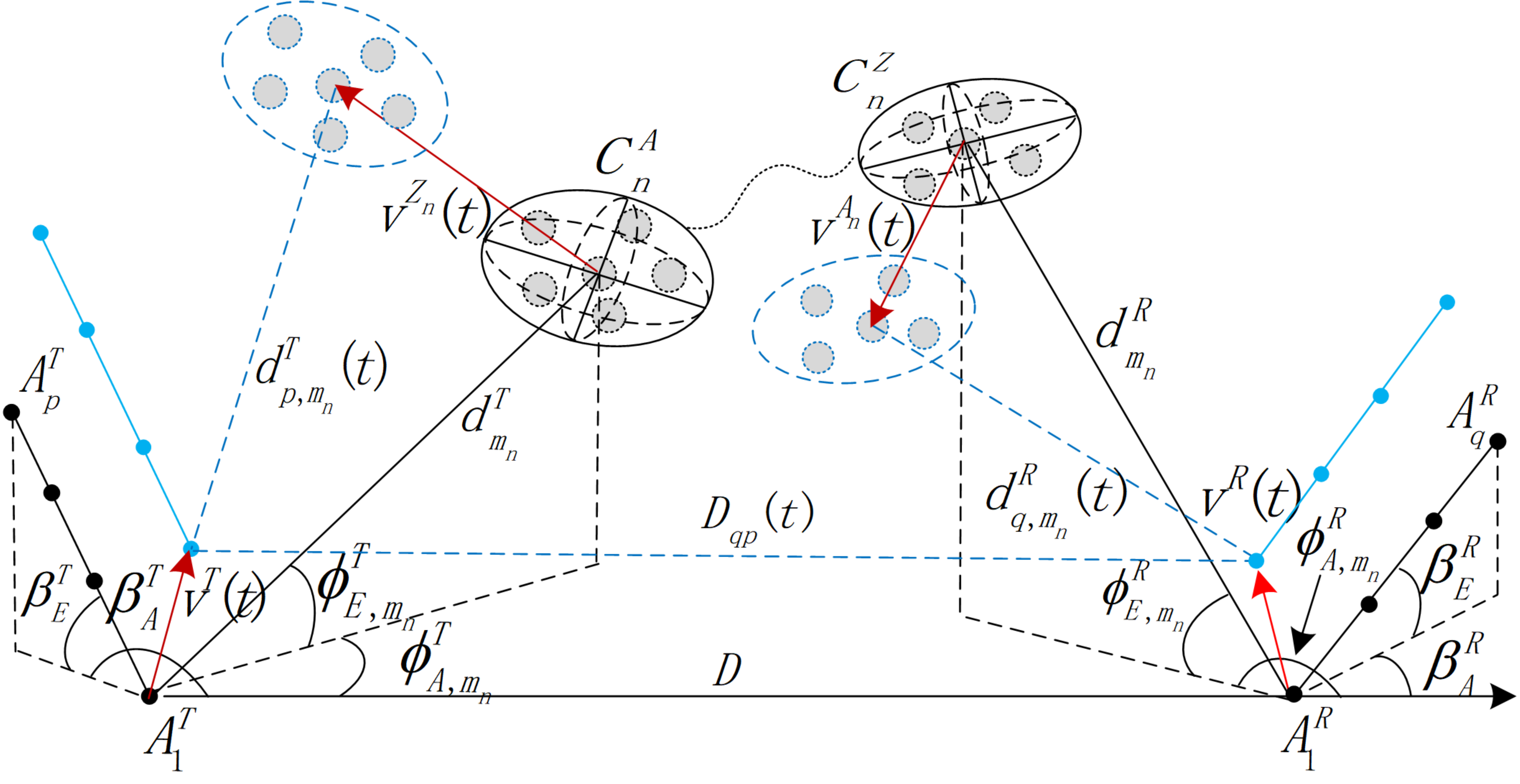}
\caption{A 3D non-stationary ultra-wideband massive MIMO GBSM.}
\label{fig_ChannelModel}
\end{figure}

Let us consider a massive MIMO communication system.
As is shown in Fig. \ref{fig_ChannelModel},
large uniform linear arrays (ULAs) with antenna spacings $\delta^T$ and $\delta^R$ are deployed at the Tx and Rx, respectively.
Symbol $\beta_A^{T(R)}$ is the tilt angle of Tx(Rx) antenna array in the $xy$ plane, $\beta_E^{T(R)}$ is the elevation angle of the Tx(Rx) antenna array relative to the $xy$ plane.
For clarity, only the $n$th ($n=1,...,N_{qp}(t)$) cluster is shown in this figure.
The $n$th path is represented by one-to-one pair clusters, i.e., $C^A_n$ at the Tx side and the cluster $C^Z_n$ at the Rx side.  $N_{qp}(t)$ is the total number of paths in the link between the $p$th ($p=1,...,M_T$) Tx antenna $A^T_p$ and the $q$th ($q=1,...,M_R$) Rx antenna $A^R_q$ at time instant $t$.
The propagation between $C^A_n$ and $C^Z_n$ is abstracted by a virtual link \cite{WINNERII}.
There could be other clusters between $C^A_n$ and $C^Z_n$, introducing more than two reflections/interactions between the Tx and Rx.
When the delay of the virtual link is zero, the multi-bounce rays reduce to single-bounce rays.
In this model, the Tx, Rx, and clusters are allowed to change their velocities and  trajectories.
The movements of the Tx, Rx, $C^A_n$, and $C^Z_n$ are described by the speeds $v^X(t)$, travel azimuth angles $\alpha_A^{X}(t)$, and travel elevation angles $\alpha_E^{X}(t)$, respectively.
The superscript $X\in\{T,R,A_n,Z_n\}$ denotes the Tx, Rx, $C_n^A$, and $C_n^Z$, respectively.
The azimuth angle of departure (AAoD) and elevation AoD (EAoD) of the $m$th ray in $C^A_n$ transmitted from $A^T_1$ are denoted by $\phi_{A,m_n}^T$ and $\phi_{E,m_n}^T$, respectively.
Similarly, $\phi_{A,m_n}^R$ and $\phi_{E,m_n}^R$ stand for the azimuth angle of arrival (AAoA) and elevation AoA (EAoA) of the $m$th ray in the $C^Z_n$ impinging on $A^R_1$, respectively.
The EAoD, EAoA, AAoD, and AAoA of the line-of-sight (LoS) path are denoted by $\phi_{E,L}^T$, $\phi_{E,L}^R$, $\phi_{A,L}^T$, and $\phi_{A,L}^R$, respectively.
The distances of $A^{T}_1$--$S^{A}_{m_n}$, $S^{Z}_{m_n}$--$A^{R}_1$, and $A^T_1$--$A^R_1$ are denoted by $d^{T}_{m_n}$, $d^R_{m_n}$, and $D$, respectively, where $S^{A(Z)}_{m_n}$ ($m_n=1,...,M_n$) is the $m$th scatterer in $C^{A(Z)}_n$.
Note that the above-mentioned departure/arrival angles and the distances $d^{T}_{m_n}$, $d^R_{m_n}$, and $D$ are the initial values at time $t_0$.
The time-variation of the proposed model is described in the remainder of this section.
The parameters in Fig. \ref{fig_ChannelModel} are defined in Table \ref{tab_parameters}.

\begin{table*}[ht]
\caption{SUMMARY OF KEY PARAMETER DEFINITIONS.}
\center
    \begin{tabular}{|c|l|}
    \hline
    $A_p^T$, $A_q^R$ & The $p$th Tx antenna element and the $q$th Rx antenna element, respectively \\
    \hline
    $\delta^T, \delta^R$ & Antenna spacings of the Tx and Rx arrays, respectively \\
    \hline
    $\beta_A^{T(R)},\beta_E^{T(R)}$ & Azimuth and elevation angles of the Tx(Rx) antenna array, respectively \\
    \hline
    $C_n^A$, $C_n^Z$ & The first- and last-bounce clusters of the $n$th path, respectively \\
    \hline
    $v^T(t),v^R(t),v^{A_n}(t),v^{Z_n}(t)$ & Speeds of the Tx, Rx, $C_n^A$, and $C_n^Z$, respectively \\
    \hline
    $\alpha_A^T(t),\alpha_A^R(t),\alpha_A^{A_n}(t),\alpha_A^{Z_n}(t)$ & Travel azimuth angles of the Tx, Rx, $C_n^A$, and $C_n^Z$, respectively \\
    \hline
    $\alpha_E^T(t),\alpha_E^R(t),\alpha_E^{A_n}(t),\alpha_E^{Z_n}(t)$ & Travel elevation angles of the Tx, Rx, $C_n^A$, and $C_n^Z$, respectively \\
    \hline
    $\phi^T_{A,m_n}$, $\phi^T_{E,m_n}$ & AAoD and EAoD of the $m$th ray in $C_n^A$ transmitted from $A^T_1$ at initial time, respectively \\
    \hline
    $\phi^R_{A,m_n}$, $\phi^R_{E,m_n}$ & AAoA and EAoA of the $m$th ray in $C_n^Z$ impinging on $A^R_1$ at initial time, respectively \\
    \hline
    $\phi^T_{A,L}$, $\phi^T_{E,L}$ & AAoD and EAoD of the LoS path transmitted from $A^T_1$ at initial time, respectively \\
    \hline
    $\phi^R_{A,L}$, $\phi^R_{E,L}$ & AAoA and EAoA of the LoS path impinging on $A^R_1$ at initial time, respectively \\
    \hline
    $d_{m_n}^T$, $d_{m_n}^R$ & Distance from $A_1^T$($A_1^R$) to $C_n^A$($C_n^Z$) via the $m$th ray at initial time \\
    \hline
    $d_{p,m_n}^T(t)$, $d_{q,m_n}^R(t)$ & Distance from $A_p^T$($A_q^R$) to $C_n^A$($C_n^Z$) via the $m$th ray at time instant $t$ \\
    \hline
    $D$ & Distance from $A_1^T$ to $A_1^R$ at initial time \\
    \hline
    $D_{qp}(t)$ & Distance from $A_p^T$ to $A_q^R$ at time instant $t$ \\
    \hline
    \end{tabular}
    \label{tab_parameters}
\end{table*}
\subsection{Channel Impulse Response}
Considering small-scale fading, path loss, shadowing, oxygen absorption, and blockage effect, the complete channel matrix is given by
$\textbf H=[PL\cdot SH \cdot BL\cdot OL]^{\frac{1}{2}}\cdot \textbf H_\text{s}$,
where $PL$ denotes the path loss.
Widely used path loss model can be found in \cite{5GCM}, which has been recommended as the path loss model for 5G systems.
$SH$ denotes the shadowing and is modeled as lognormal random variables \cite{5GCM}.
The blockage loss $BL$ caused by humans and vehicles is taken from \cite{METISD1.4}.
The oxygen absorption loss $OL$ for mmWave and THz communications can be found in \cite{IMT-2020} and \cite{Barros2017THz}, respectively.
Note that $PL$, $SH$, $BL$, and $OL$ are in power level, which can be transformed into the corresponding dB values as $10\log_{10}(\alpha)$, where $\alpha\in\{PL, SH, BL, OL\}$.

\begin{figure}
\centering\includegraphics[width=2.5in]{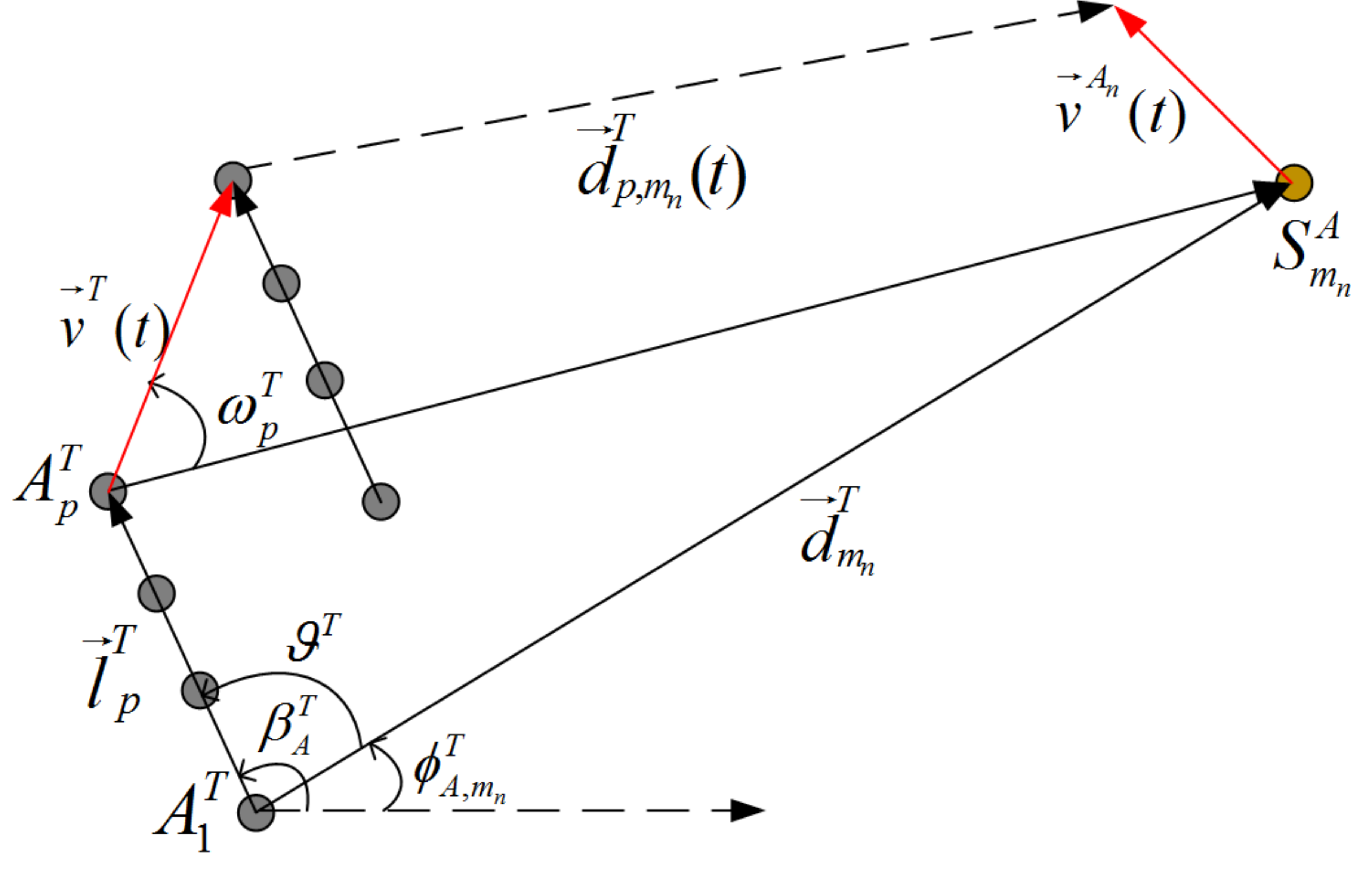}
\caption{The projection of the propagation between the Tx and $S^A_{m_n}$ on the $xy$ plane.}
\label{fig_2Dvec}
\end{figure}

The small-scale fading is represented as a complex matrix $\textbf H_s=[h_{qp}(t,\tau)]_{M_R\times M_T}$, where $h_{qp}(t,\tau)$ is the CIR between $A^T_p$ and $A^R_q$ and expressed as the summation of the LoS and non-LoS (NLoS) components, i.e.,
\begin{equation}\label{CIR}
h_{qp}(t,\tau)=\sqrt{\frac{K_R}{K_R+1}}h_{qp}^L(t,\tau)+\sqrt{\frac{1}{K_R+1}}h_{qp}^N(t,\tau)
\end{equation}
where $K_R$ is the K-factor.
The NLoS components $h_{qp}^N(t,\tau)$ can be written as
\begin{align} \label{CIR_N}
&h_{qp}^N(t,\tau)= \sum_{n=1}^{N_{qp}(t)}\sum_{m=1}^{M_n}
\begin{bmatrix}
    F_{q,V}(\phi^R_{E,m_n},\phi^R_{A,m_n})\\
    F_{q,H}(\phi^R_{E,m_n},\phi^R_{A,m_n})
    \end{bmatrix}^\text{T}\nonumber \\ &\cdot \begin{bmatrix}
    e^{j\theta_{m_n}^{VV}} & \sqrt{\mu\kappa^{-1}_{m_n}}e^{j\theta_{m_n}^{VH}}\\
    \sqrt{\kappa^{-1}_{m_n}}e^{j\theta_{m_n}^{HV}} & \sqrt{\mu}e^{j\theta_{m_n}^{HH}}
    \end{bmatrix}
    \begin{bmatrix}
    F_{p,V}(\phi^T_{E,m_n},\phi^T_{A,m_n})\\
    F_{p,H}(\phi^T_{E,m_n},\phi^T_{A,m_n})
    \end{bmatrix}  \nonumber\\
&\cdot \sqrt{P_{qp,m_n}(t)}e^ {j2\pi f_c \tau_{qp,m_n}(t)}  \cdot\delta(\tau-\tau_{qp,m_n}(t))
\end{align}
where $\{\cdot \}^\text T$ denotes transposition,
$f_c$ is the carrier frequency,
$F_{p(q),V}$ and $F_{p(q),H}$ are the antenna patterns of $A_p^T$($A_q^R$) for vertical and horizontal polarizations, respectively.
Note that the proposed propagation channel model is designed to be antenna independent, which means different antenna patterns can be applied without modifying the basic model framework.
Symbol $\kappa_{m_n}$ stands for the cross polarization power ratio \cite{3GPP38901}, $\mu$ is co-polar imbalance, $\theta_{m_n}^{VV}$, $\theta_{m_n}^{VH}$, $\theta_{m_n}^{HV}$, and $\theta_{m_n}^{HH}$ are initial phases with uniform distribution over $(0,2\pi]$,
$P_{qp,m_n}(t)$ and $\tau_{qp,m_n}(t)$ are the powers and delays of the $m$th ray in the $n$th cluster between $A^T_p$ and $A^R_q$ at time $t$, respectively.
Considering large sizes of antenna array and high-mobility scenarios, the non-stationarities on time axis and array axis have to be considered.
The number of clusters $N_{qp}(t)$, the power of ray $P_{qp,m_n}(t)$, and the propagation delay $\tau_{qp,m_n}(t)$ are modeled as space and time-dependent.
The propagation delay $\tau_{qp,m_n}(t)$ is determined as
\begin{equation}\label{equ_Delay}
\tau_{qp,m_n}(t)={d_{qp,m_n}(t)}/{c}+\tilde{\tau}_{m_n}.
\end{equation}
Here, $c$ denotes the speed of light, $\tilde{\tau}_{m_n}$ indicates the delay of the link between $S_{m_n}^A$ and $S_{m_n}^Z$, and is modeled as $\tilde \tau_{m_n}=\tilde d_{m_n}/c+\tau_{C,\text{link}}$, where $\tilde d_{m_n}$ is the distance of $S_{m_n}^A$--$S_{m_n}^Z$, $\tau_{C,\text{link}}$ is a non-negative variable randomly generated according to exponential distribution \cite{Verdone2012Pervasive}.
The travel distance $d_{qp,m_n}(t)$ is expressed as
$ d_{qp,m_n}(t)= \|\vec d^T_{p,m_n}(t)\|+ \|\vec d^R_{m_n,q}(t)\| $,
where $\|\cdot\|$ stands for the Frobenius norm,
$\vec d^T_{p,m_n}(t)$ and $\vec d^R_{m_n,q}(t)$ are the vector from $A^T_p$ to $S^A_{m_n}$ and the vector from $A^R_q$ to $S^Z_{m_n}$ at time $t$, respectively.
Since the symmetry of the propagation, only the first-bounce propagation between $A^T_p$ and $S^A_{m_n}$ is described.
For the sake of clarity, Fig. \ref{fig_2Dvec} shows the projection of propagation between the Tx and $S^A_{m_n}$ on the $xy$ plane.
Considering the time-varying speeds and trajectories of the Tx and $C_n^A$, $\vec d^T_{p,m_n}(t)$ is calculated as
\begin{equation}\label{equ_vecT}
\vec d^T_{p,m_n}(t)=\vec d^T_{m_n}-[\vec l^T_p+\int_{0}^{t} \vec v^{T}(t)-\vec v^{A_n}(t)\rm dt]
\end{equation}
where
\begin{align}
\vec d^T_{m_n}=d^T_{m_n}\begin{bmatrix}
\cos(\phi^T_{E,m_n})\cos(\phi^T_{A,m_n})\\
\cos(\phi^T_{E,m_n})\sin(\phi^T_{A,m_n})\\
\sin(\phi^T_{E,m_n})
\end{bmatrix}^\text T
\end{align}
\begin{equation}\label{equ_antenna}
\vec l^T_{p}=\delta_p\begin{bmatrix}
\cos(\beta^T_{E})\cos(\beta^T_{A})\\
\cos(\beta^T_{E})\sin(\beta^T_{A})\\
\sin(\beta^T_{E})]
\end{bmatrix} ^{\text T}
\end{equation}
\begin{equation}
\vec v^T(t)=v^T(t)\begin{bmatrix}
\cos\left(\alpha^T_{E}(t)\right)\cos\left(\alpha^T_{A}(t)\right)\\
\cos\left(\alpha^T_{E}(t)\right)\sin(\alpha^T_{A}(t))\\
\sin\left(\alpha^T_{E}(t)\right)
\end{bmatrix}^\text T
\end{equation}
\begin{equation}
\vec v^{A_n}(t)=v^{A_n}(t)\begin{bmatrix}
\cos(\alpha^{A_n}_{E}(t))\cos(\alpha^{A_n}_{A}(t))\\
\cos(\alpha^{A_n}_{E}(t))\sin(\alpha^{A_n}_{A}(t))\\
\sin(\alpha^{A_n}_{E}(t))]
\end{bmatrix}^\text T.
\end{equation}
Here $\delta_p=(p-1)\delta^T$, indicates the distance of $A^T_p$--$A^T_1$.

For most cases, the Tx, Rx, and scatterers move in the $xy$ plane.
For conciseness, we use $v^T=\|\vec v^{T}-\vec v^{A_n}\|$ and $\alpha^T=\text{arg}\{\vec v^{T}-\vec v^{A_n}\}$ to denote the relative speed and angle of motion of the Tx w.r.t. $C_n^A$, respectively, where $\text{arg}\{\cdot\}$ calculates the argument of a 2D vector.
When the Tx, Rx and clusters move with constant speeds along straight trajectories, by extending the 2D parabolic wavefront \cite{Lopez2016} into 3D time non-stationary case, travel distance $d^T_{p,m_n}(t)=\|\vec d^T_{p,m_n}(t) \|$ is approximated as
\begin{align}\label{TraDis1}
d^T_{p,m_n}(t)&\approx \underbrace{d^T_{m_n}-\cos(\omega^T_p) v^T t-\cos(\vartheta^T)\delta_p}_{\text{WSS PWF approximation}}\nonumber\\ &\underbrace{+\frac{\sin^2(\vartheta^T)\delta_p^2}{2d^T_{m_n}}}_{\text{SWF term}}+\underbrace{\frac{\sin^2(\omega_p^T)(v^T t)^2}{2[d^T_{m_n}-\cos(\vartheta^T)\delta_p]}}_\text{non-WSS term}
\end{align}
where  $\vartheta^T$ is the angle between the the transmit antenna array and the $m$th ray in the $n$th cluster transmitted from $A^T_1$, and is calculated as
\begin{align}\label{theta}
\cos(\vartheta^T)&=\cos(\phi^T_{E,m_n})\cos(\beta^T_{E})\cos(\beta^T_A-\phi^T_{A,m_n}) \nonumber \\ & + \sin(\phi^T_{E,m_n})\sin(\beta^T_E).
\end{align}
In (\ref{TraDis1}), $\omega_p^T$ stands for the angle from moving direction of the Tx to the $m$th ray of the $n$th cluster transmitted from $A^T_p$, and can be determined as
\begin{small}
\begin{align}\label{omegap}
&\cos(\omega_p^T)= \nonumber \\ &\frac{d^T_{m_n}\cos(\alpha^T-\phi^T_{A,m_n})\cos(\phi^T_{E,m_n})-\delta_p\cos(\alpha^T-\beta^T_A)\cos(\beta^T_E)}{[(d^T_{m_n})^2-2d^T_{m_n}\delta_p\cos(\vartheta^T)+\delta_p^2]^{1/2}}.
\end{align}
\end{small}

Equation (\ref{TraDis1}) gives an efficient and scalable approach for modeling the 3D SWF under time non-stationary assumption.
The first term in \eqref{TraDis1} gives the travel distance of $A_p^T$--$S^A_{m_n}$ link based on plane wavefront (PWF) and temporal WSS assumptions.
The second and third terms account for the non-stationary properties of the channel in the space and time domains, respectively.
Under certain conditions, the travel distance in \eqref{TraDis1} can be further simplified.

\subsubsection{Case I: non-WSS \& PWF}
When small antenna arrays are used,  i.e., $\delta_p \ll d^T_{m_n}$,
the angle $\omega_p^T$ becomes constant for different antenna elements and \eqref{omegap} reduces to
\begin{equation}\label{omega}
\cos(\omega^T)=\cos(\alpha^T-\phi^T_{A,m})\cos(\phi^T_{E,m}).
\end{equation}
Note that the subscript ``$p$" has been omitted for convenience.
The SWF term in \eqref{TraDis1} tends to zero and $d^T_{p,m_n}(t)$ reduces to
\begin{align}
d^T_{p,m_n}(t)&\approx d^T_{m_n}-\cos(\omega^T) v^T t-\cos(\vartheta^T) \delta_p\nonumber \\ &+ \frac{\sin^2(\omega^T)( v^T t)^2}{2[d^T_{m_n}-\cos(\vartheta^T) \delta_p]}.
\end{align}
\subsubsection{Case II: WSS \& SWF}
For slow-moving scenarios or short time periods, i.e. $v^T t \ll d^T_{m_n}$, the non-WSS term in \eqref{TraDis1} tends to zero, which makes the model stationary over the time.
The travel distance in \eqref{TraDis1} reduces to
\begin{equation} \label{caseII}
d^T_{p,m_n}(t)\approx d^T_{m_n}-\cos(\omega^T_p) v^T t-\cos(\vartheta^T) \delta_p+\frac{\sin^2(\vartheta^T)\delta_p^2}{2d^T_{m_n}}.
\end{equation}
\subsubsection{Case III: WSS \& PWF}
When both time WSS and PWF conditions are fulfilled. The angle $\omega^T_p$ is simplified according to (\ref{omega}).
The travel distance in (\ref{TraDis1}) reduces to a fundamental expression, which can be found in most of the existing channel models as \cite{IMTA,WINNERII,WINNERPlus,3GPP3D}
\begin{equation} \label{TraDis4}
d^T_{p,m_n}(t)\approx d^T_{m_n}-\cos(\omega^T) v^T t-\cos(\vartheta^T) \delta_p.
\end{equation}

For the Rx side, $\vec d^R_{m_n,q}(t)$, $\vartheta^R$, and $\omega_q^R$ are obtained by replacing the superscript ``$T$" and subscript ``$p$" with ``$R$" and ``$q$" in (\ref{equ_vecT})--(\ref{TraDis4}), respectively.
Here, we briefly discuss the influence of the Doppler shifts on the proposed model.
In \eqref{CIR_N} the phase rotation associated with time-varying travel distance is given as $\varphi_{qp,m_n}(t)=2\pi f_c\tau_{qp,m_n}(t)$, and is decomposed as $\varphi_{qp,m_n}(t)=2\pi f_c( \|\vec d_{p,m_n}^T(t)+\vec d^R_{q,m_n}(t)\|/c+\tilde \tau_{m_n})$.
The Doppler shift can be estimated by $f_{m_n}(t)= \frac{\mathrm d  \varphi_{qp,m_n}(t)}{\mathrm d t}$ and is time-varying.
Considering the WSS\ \&\ SWF case in \eqref{caseII}, the phase rotation can be further expressed as $\varphi_{qp,m_n}(t)=\frac{2\pi}{\lambda}\left(d_{p,m_n}^T+\tilde \tau_{m_n}c+d_{m_n,q}^T \right )-2\pi t(f^T_{m_n}+f^R_{m_n})$, where $d_{p,m_n}^T+\tilde \tau_{m_n}c+d_{m_n,q}^T$ accounts for the distance of $A^T_p$--$S_{m_n}^A$--$S_{m_n}^Z$--$A_q^R$ link, $f_{m_n}^T=\frac{v^T}{\lambda}\cos\omega_p^T$ and $f_{m_n}^R=\frac{v^R}{\lambda}\cos\omega_q^R$ are the Doppler shifts caused by the movement of the Tx relative to $C^A_n$ and the movement of the Rx relative to $C^Z_n$, respectively.
Finally, the LoS component in (\ref{CIR}) is calculated as
\begin{align}  \label{CIR_L}
&h_{qp}^{L}(t,\tau)=
\begin{bmatrix}
    F_{q,V}(\phi^R_{E,L},\phi^R_{A,L})\\
    F_{q,H}(\phi^R_{E,L},\phi^R_{A,L})
    \end{bmatrix}^\text{T}
    \cdot \begin{bmatrix}
    e^{j\theta_{L}^{VV}} & 0\\
    0 & -e^{j\theta_{L}^{HH}}
    \end{bmatrix}\nonumber \\
    &\cdot \begin{bmatrix}
    F_{p,V}(\phi^T_{E,L},\phi^T_{A,L})\\
    F_{p,H}(\phi^T_{E,L},\phi^T_{A,L})
    \end{bmatrix}\exp \{j[2\pi f_c \tau_{qp}^{L}(t)]\} \delta(\tau-\tau_{qp}^{L}(t))
\end{align}
where $\theta_L^{VV}$ and $\theta_L^{HH}$ are random phases with uniform distribution over $(0,2\pi]$, $\tau_{qp}^L(t)$ are space and time-variant propagation delay of the LoS path, and determined as $\tau_{qp}^L(t)=D_{qp}(t)/c$, where $D_{qp}(t)=\|\vec D_{qp}(t)\|$ is the distance between $A^T_p$ and $A^R_q$. The vector $\vec D_{qp}(t)$ is calculated as
\begin{equation}
\vec D_{qp}(t)=\vec D+\vec l_q^R-\vec l_p^T+\int_0^t \vec v^R(t)-\vec v^T(t) \rm d t
\end{equation}
where $\vec D=[D,0,0]$. When the Tx and Rx travel in the
horizontal plane with constant velocity, $D_{qp}(t)$ can be determined as
\begin{align}
&[D_{qp}(t)]^2=[D+\cos(\alpha^R)v^Rt-\cos(\alpha^T)v^Tt \nonumber \\ & -\cos(\beta^T_A)\cos(\beta^T_E)\delta_p+\cos(\beta^R_A)\cos(\beta^R_E)\delta_q]^2 \nonumber \\ & +[\sin(\alpha^R)v^Rt-\sin(\alpha^T)v^Tt -\cos(\beta^T_E)\sin(\beta^T_A)\delta_p \nonumber\\ & +\cos(\beta_E^R)\sin(\beta_A^R)\delta_q]^2+[\sin(\beta_E^T)\delta_p-\sin(\beta_E^R)\delta_q]^2.
\end{align}
The space and time-varying transfer function $H_{qp}(t,f)$ is calculated as the Fourier transform of $h_{qp}(t,\tau)$ w.r.t. $\tau$, i.e.,
\begin{equation}\label{equ_TF}
H_{qp}(t,f)=\sqrt{\frac{K_R}{K_R+1}}H^L_{qp}(t,f)+\sqrt{\frac{1}{K_R+1}}H^N_{qp}(t,f).
\end{equation}
For simplicity, we use omnidirectional antennas and consider vertical polarization.
The LoS and NLoS components of the transfer function are written as
\begin{equation}\label{equ_TFLoS}
H_{qp}^L(t,f)= \exp \{j2\pi \tau_{qp}^L(t)(f_c-f)\}
\end{equation}
\begin{align}\label{equ_TFNLoS}
H_{qp}^N(t,f)&= \sum_{n=1}^{N_{qp}(t)}\sum_{m=1}^{M_n} \sqrt{P_{qp,m_n}(t)}\nonumber \\ &\times \exp \{j[\theta_{m_n}+2\pi \tau_{qp,m_n}(t)(f_c-f)]\}.
\end{align}

For the case when the system bandwidth is relatively large, e.g., $B/f_c>20\%$, the frequency dependence of the channel cannot be neglected and the US assumption may not be fulfilled \cite{Molisch2005UWB}.
A typical approach for the non-US assumption is to model the path gain as frequency-dependent \cite{He2017THz}.
The NLoS components of the space and time-varying transfer function is rewritten as
\begin{align}\label{equ_TFNLoS2}
H_{qp}^N(t,f)=&\sum_{n=1}^{N_{qp}(t)} \sum_{m=1}^{M_n}\sqrt{P_{qp,m_n}(t)} (\frac{f}{f_c})^{\gamma_{m_n}}\nonumber \\ \times &\exp \{j[\theta_{m_n}+2\pi \tau_{qp,m_n}(t)(f_c-f)]\}
\end{align}
where $\gamma_{m_n}$ is a environment-dependent random variable.

\subsection{Space and Time-Varying Ray Power}
For most of the standard 5G channel models, e.g., \cite{Wu2017unified}, \cite{3GPP38901}, and \cite{IMT-2020}, the cluster powers are constant for different antenna elements, which may be inconsistent with the measurement results~\cite{Gao2015MaMIMO}.
Based on \cite{3GPP38901}, the ray power between $A^T_p$ and $A^R_q$ at time $t$ is given as
\begin{equation} \label{equ_Power1}
P'_{qp,m_n}(t)=\exp\left( -\tau_{qp,m_n}(t)\frac{r_\tau-1}{r_\tau DS} \right )10^{\frac{-Z_n}{10}}\cdot \xi_{n}(q,p)
\end{equation}
where $Z_n$ is the per cluster shadowing term in dB, $DS$ is the root mean square (RMS) delay spread, $r_\tau$ denotes the delay distribution proportionality factor and determined as the ratio of the standard deviation of the delays to the RMS delay spread \cite{3GPP38901}.
The smooth power variations over the transmit and receive arrays can be simulated by a 2D spatial lognormal process $\xi_{n}(q,p)$, and can be calculated as
$\xi_{n}(q,p)=10^{[\mu_{n}(q,p)+\sigma_{n}\cdot s_n(q,p)]/10}$
where $\mu_n(q,p)$ is the local mean and $s_n(q,p)$ is a 2D Gaussian process, which account for the path loss and shadowing along the large arrays, respectively.
The final ray powers are obtained by normalizing $P'_{qp,n_m}(t)$ as
\begin{equation}\label{equ_normalized}
P_{qp,m_n}(t)=P'_{qp,m_n}(t)\big/\sum_{n=1}^{N_{qp}(t)}\sum_{m_n=1}^{M_n} P'_{qp,m_n}(t).
\end{equation}

The space and time-varying ray power can be simplified under certain condition.
For example, when conventional antenna array is employed at the Rx, the 2D spatial lognormal process $\xi_{n}(q,p)$ reduces to one-dimensional (1D) process $\xi_{n}(p)$.
Imposing $\xi_{n}(q,p)=1$ indicates farfield condition is fulfilled at both ends.
Furthermore, if the delays within a cluster are unresolvable, the cluster power can be generated by replacing $\tau_{qp,n_m}(t)$ with cluster delay $\tau_{qp,n}(t)$, where $\tau_{qp,n}(t)=[\sum_{m_n=1}^{M_n}\tau_{qp,m_n}(t)]/M_n$.
The ray powers within a cluster are equally determined as
\begin{equation}
P'_{qp,m_n}(t)=\frac{1}{M_n}\exp\left( -\tau_{qp,n}(t)\frac{r_\tau-1}{r_\tau DS} \right )10^{\frac{-Z_n}{10}}\cdot \xi_{n}(q,p).
\end{equation}
Finally, the ray powers are normalized as (\ref{equ_normalized}).
\subsection{Unified  Space-Time Evolution of Clusters}
Channel measurements have shown that in high-mobility scenarios, e.g., V2V and HST scenarios, clusters exhibit a birth-death behavior over time \cite{Yang2019V2V}.
In massive MIMO communication systems, similar properties can be observed on the array axis  \cite{Li2019MaMIMO}.
Here, the space-time cluster evolutions are modeled in a uniform manner.
For the Tx side, the probability of a cluster remains over time interval $\Delta t$ and antenna element spacing $\delta_p$ can be calculated as
\begin{align}\label{equ_Premain}
&P^T_\text{remain}(\Delta t,\delta_p)\nonumber \\ &= \exp\left( {-\lambda_R[( \varepsilon_1^T)^2+(\varepsilon _2^T)^2-2\varepsilon _1^T\varepsilon_2^T\cos(\alpha_A^T-\beta_A^T)]^{\frac{1}{2}}}\right ).
\end{align}
The process is described by the cluster generation rate $\lambda_G$ and cluster recombination (disappearance) rate $\lambda_R$, which can be estimated as in \cite{Saito2011_BirthDeath}.
Note that $\lambda_G$ and $\lambda_R$ are related to characteristics of scenarios and antenna patterns.
In \eqref{equ_Premain}, $\varepsilon_1^T=\delta_p\cos(\beta_E^T)/D_c^A$ and $\varepsilon_ 2^T=v^T\Delta t/D_c^S$ characterize the position differences of transmit antenna element on array and time axes, respectively.
Symbols $D_c^A$ and $D_c^S$ are scenario-dependent correlation factors in the array and time domains, respectively.
Typical values of $D_c^A$ and $D_c^S$ such as 10~m and 30~m can be chosen, which are the same order of correlation distances in \cite{Wu2017unified,3GPP38901}.

For the Rx side, the probability of a cluster exist over time interval $\Delta t$ and element spacing $\delta_q$, i.e., $P^R_\text{remain}(\Delta t,\delta_q)$, is calculated similarly.
Since each antenna element has its own observable cluster set, only a cluster can be seen by at least one Tx antenna and one Rx antenna, it can contribute to the received power.
Therefore, the joint probability of a cluster exist over $\Delta t$ and $\delta_q$ is calculated as
\begin{equation}
P_\text{remain}(\Delta t, \delta_p,\delta_q)=P^T_\text{remain}(\Delta t,\delta_p)\cdot P^R_\text{remain}(\Delta t,\delta_q).
\end{equation}
The mean number of newly generated clusters is obtained by
\begin{equation}
\mathbb{E}\{N_\text{new}\}=\frac{\lambda_G}{\lambda_R}[1-P_\text{remain}(\Delta t, \delta_p,\delta_q)].
\end{equation}

\subsection{Ellipsoid Gaussian Scattering Distribution}
The Gaussian scatter density model (GSDM) has widely been used in channel modeling for various communication scenarios and validated by the measurement data \cite{Wong2010_DoA,Andrade2003GSDM}.
In GSDM, the scatterers are gathered around their center and usually modeled by certain shapes, e.g., discs in 2D models and spheres in 3D models \cite{Mammasis2012Gaussian}.
However, channel measurements indicate that the spatial dispersions of scatterers within a cluster, which can be described by cluster angular spread (CAS), cluster elevation spread (CES), and cluster delay spread (CDS), are usually unequal \cite{METISD1.4, WINNERPlus}.
Based on the aforementioned assumption, the positions of scatterers centering on the origin of coordinates are modeled as
\begin{equation}
p(x',y',z')=\frac{\exp\left( -\frac{x'^2}{2\sigma_{DS}^2}-\frac{y'^2}{2\sigma_{AS}^2}-\frac{z'^2}{2\sigma_{ES}^2}\right )}{{(2\pi)}^{3/2}\sigma_{DS}\sigma_{AS}\sigma_{ES}}
\end{equation}
where $\sigma_{DS}$, $\sigma_{AS}$, and $\sigma_{ES}$ are the standard derivations of the Gaussian distributions and characterize the CDS, CAS, and CES, respectively.
The scatterers centering around the spherical coordinates $(\overline d, \overline \phi_E, \overline \phi_A)$ can be obtained by shifting the above-mentioned cluster using the following transformation
\begin{align} \label{equ_rotate}
\begin{bmatrix}
x\\y\\z
\end{bmatrix}&=\begin{bmatrix}
\cos(\overline \phi_A) &-\sin(\overline \phi_A)  & 0\\
\sin(\overline \phi_A)& \cos(\overline \phi_A) & 0 \\
0 & 0 & 1
\end{bmatrix} \nonumber \\
&\cdot \begin{bmatrix}
\cos(\overline \phi_E) &0  & -\sin(\overline \phi_E)\\
0& 1 & 0 \\
\sin(\overline \phi_E) & 0 & \cos(\overline \phi_E)
\end{bmatrix} \cdot
\begin{bmatrix}
x'-\overline d\\y'\\z'
\end{bmatrix}
\end{align}
where $\overline d$ denotes the distance from the Tx/Rx to the center of the cluster, $\overline \phi_E$ and $\overline \phi_A$ are the mean values of the elevation angles and azimuth angles of scatterers, respectively.
Note that the orientation of the cluster toward the Tx/Rx is constant through the aforementioned transformation, which ensures the values of CDS, CAS, and CES remain unchanged.
By substituting $x=d\cos(\phi_E) \cos(\phi_A)$, $y=d\cos(\phi_E) \sin(\phi_A)$, and $z=d\sin(\phi_E)$ in to (\ref{equ_rotate}), after some manipulations, the angle distance joint distribution can be obtained as
\begin{align}\label{equ_PDF}
&p(d,\phi_E,\phi_A)=|J(x',y',z')|\cdot p(x',y',z')\bigg|\nonumber \\
&\begin{small}
\begin{matrix}
x'=d[\cos(\phi_E)\cos(\overline \phi_E) \cos(\phi_A-\overline \phi_A)+\sin(\phi_E)\sin(\overline\phi_E)]-\overline d
 \\
\hspace{-4.4cm}y'=d\cos(\phi_E)\sin(\phi_A-\overline\phi_A)
 \\
\hspace{-0.5cm}z'=d[\sin(\phi_E)\cos(\overline\phi_E)-\cos(\phi_A-\overline \phi_A)\cos(\phi_E)\sin(\overline \phi_E)]
\vspace{-0.3cm}
\end{matrix}
\end{small}
\end{align}

where
\begin{equation}
J(x',y',z')=\begin{vmatrix}
\frac{\partial x'}{\partial d} &\frac{\partial x'}{\partial \phi_E}  &\frac{\partial x'}{\partial \phi_A} \\
\frac{\partial y'}{\partial d} &\frac{\partial y'}{\partial \phi_E}  &\frac{\partial y'}{\partial \phi_A} \\
\frac{\partial z'}{\partial d} &\frac{\partial z'}{\partial \phi_E}  &\frac{\partial z'}{\partial \phi_A}
\end{vmatrix}=-d^2\cos(\phi_E).
\end{equation}


By imposing $\tilde \tau_{m_n}=0$ in \eqref{equ_Delay}, the multi-bounce rays reduce to single-bounce rays.
The delay angle joint distribution for the cluster can be obtained based on (\ref{equ_PDF}) by transform parameter $d$ into $\tau$, i.e.,
\begin{equation}
p(\tau,\phi_E^X,\phi_A^X)=|J(d^X)|\cdot p(d^X,\phi_E^X,\phi_A^X)
\end{equation}
where
\begin{equation}
d^X=\frac{(\tau c)^2-D^2}{2\tau c-2D\cos(\phi_E^X)\cos(\phi_A^X)}
\end{equation}
\begin{equation}
J(d^X)=\frac{\mathrm{d} d^X}{\mathrm{d}\tau}=\frac{c[D^2+(\tau c)^2-2Dc\tau\cos(\phi_E^X)\cos(\phi_A^X)]}{2[\tau c-D\cos(\phi_E^X)\cos(\phi_A^X)]^2}
\end{equation}
and $X\in\{T,R\}$.
The angular parameters and travel distances of the first- and last-bounce propagations becomes interdependent.
The relationship between them can be expressed as
\begin{equation}
\tan(\phi^{T}_{A})=\frac{d^{R}\cos(\phi^R_{E})\sin(\phi^{R}_{A})}{D+d^{R}\cos(\phi_E^R)\cos(\phi^{R}_{A})}
\end{equation}
\begin{align}
&\tan(\phi^{T}_{E})=\nonumber \\ &\frac{d^R\sin(\phi_{E}^R)}{\{D^2+[d^R\cos(\phi^R_{E})]^2+2D\cdot d^R\cos(\phi_{E}^R)\cos(\phi_{A}^R)\}^{1/2}}
\end{align}
\begin{equation}
d^T=[D^2+(d^R)^2+2D\cdot d^R\cos(\phi_{E}^R)\sin(\phi_{A}^R)]^{1/2}.
\end{equation}
Note that the subscripts $m_n$ are omitted for clarity.
Unlike the WINNER/3GPP channel models \cite{WINNERPlus,3GPP3D, 3GPP38901}, where the clusters are randomly generated for every link, in the proposed model, the clusters are assumed to physically exist in the propagation environments.
Thus, spatial consistency can be achieved based on the locations of clusters.
This makes it possible to prediction channel state information (CSI) based on the channel associated with nearby users or in the previous time snapshots.
For instance, in HST scenarios, different trains travel to the same location of the track will see similar environments and hence have similar channel behaviors.
Channel can be estimated from the communication process of last trains or nearby remote radio heads (RRHs) \cite{Wang2015CSI}.

\begin{figure}
\begin{minipage}[t]{0.5\linewidth}
\centering
\includegraphics[width=1.8in]{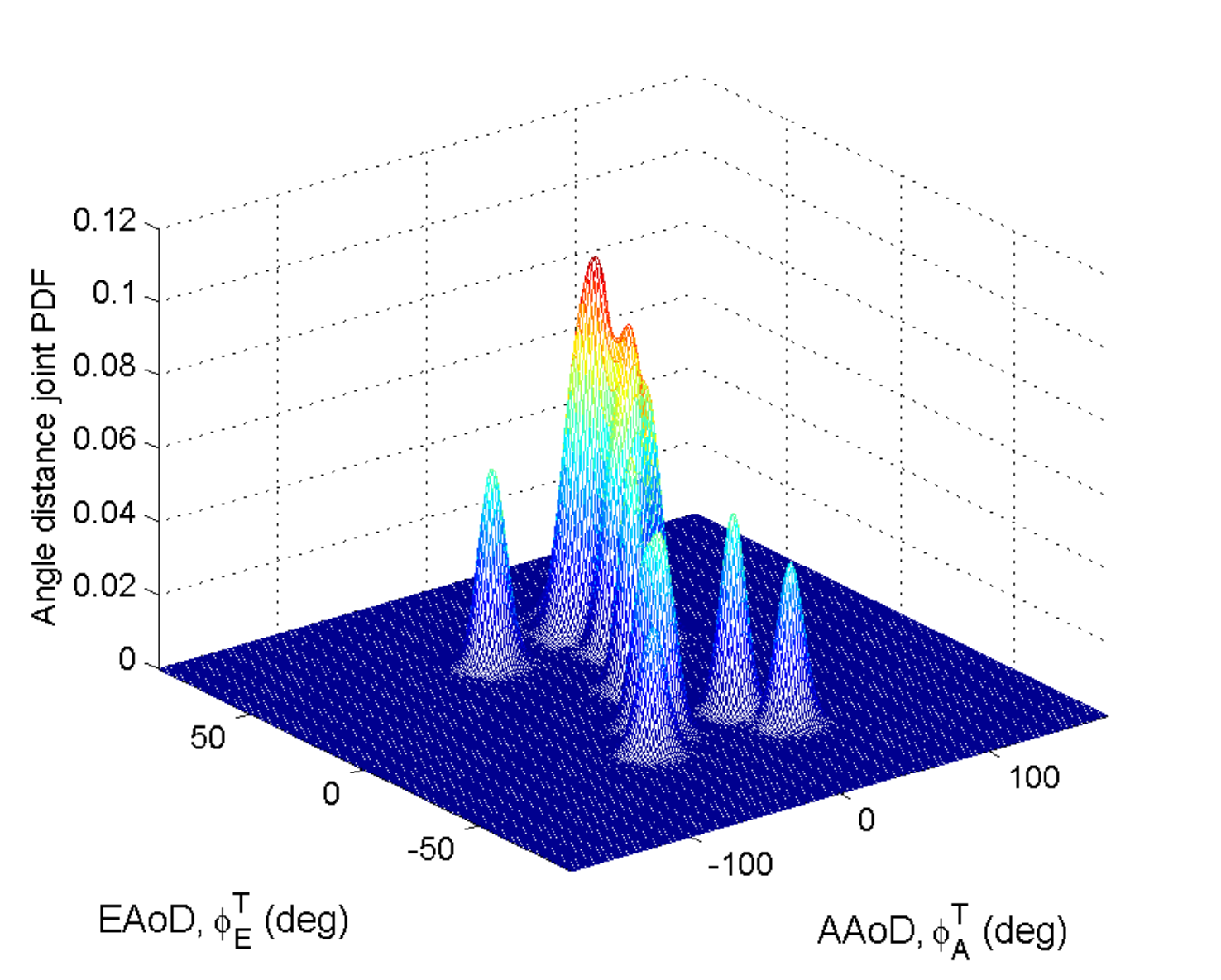}
\centerline{(a)}\tiny
\end{minipage}%
\begin{minipage}[t]{0.5\linewidth}
\centering
\includegraphics[width=1.8in]{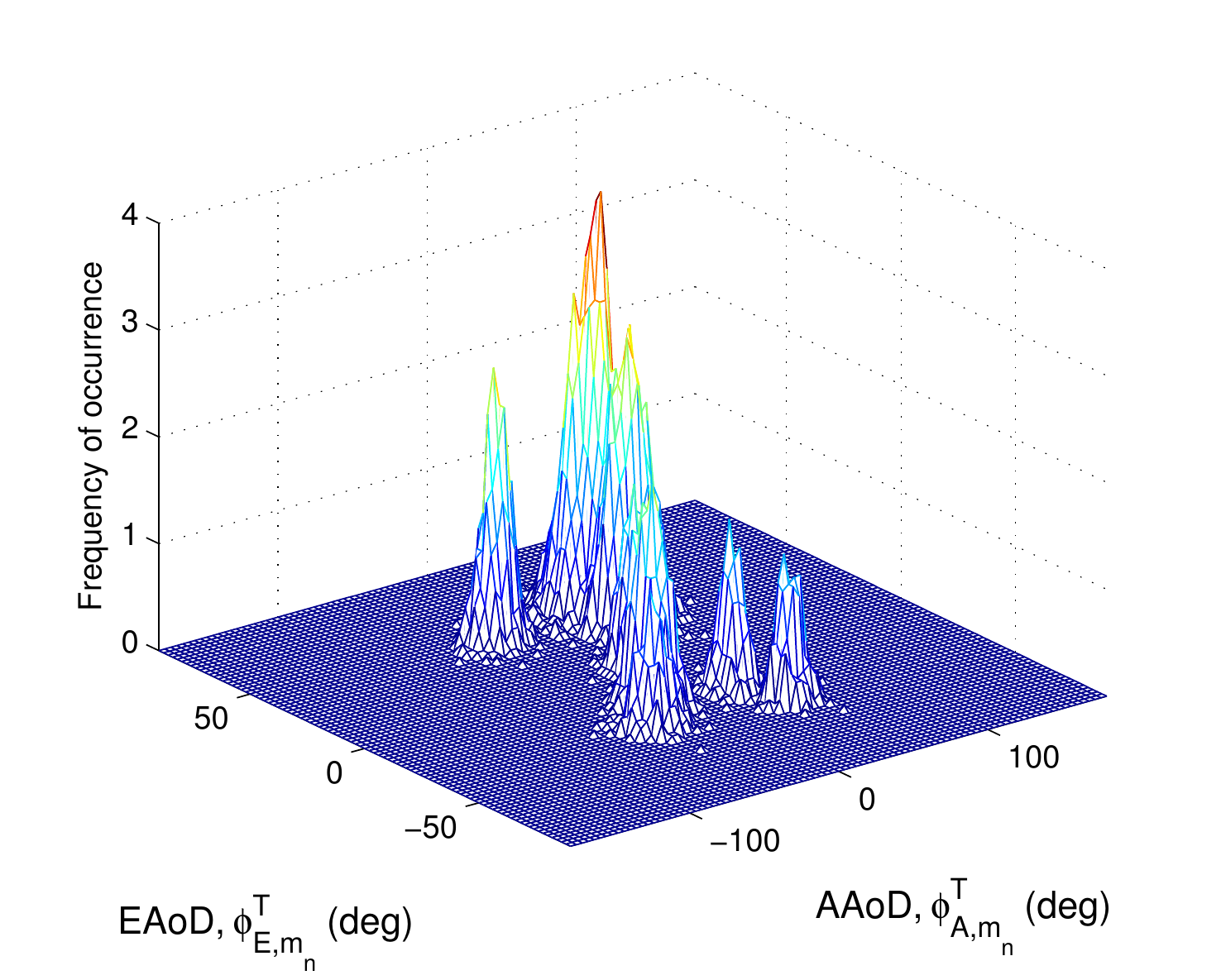}
\centerline{(b)}\tiny
\end{minipage}
\caption{(a) Theoretical and (b) simulated ellipsoid Gaussian scattering distribution ($\sigma_{DS}=8$, $\sigma_{AS}=10$, $\sigma_{ES}=6$, $\overline d \sim \mathcal N(100,10)$ m).}
\label{fig_PDF}
\end{figure}

Fig. \ref{fig_PDF} shows the theoretical and simulated ellipsoid Gaussian scattering distribution.
The mean angles, i.e., $\overline\phi_E$ and $\overline\phi_A$ are obtained from \cite{3GPP38901} in urban micro-cell scenario, NLoS case.
The distances between the Tx and the center of the first-bounce cluster follows a Gaussian distribution, i.e., $\mathcal N(100,10)$~m.
The simulated result is obtained using Monte Carlo method and a total of 500 rays are generated.
A good consistency between theoretical and simulated results can be observed.

By adjusting the model parameters or components, the proposed model can be applied to various scenarios.
Let's take mmWave-THz scenario as an example.
The path loss, shadowing, oxygen absorption, and blockage attenuation components can be replaced with those at mmWave-THz bands.
The sparsity of MPCs can be represented by adjusting the number of clusters and the number of rays within a cluster.
The remarkable birth-death behaviour of clusters over time resulting from large propagation loss can be modeled by increasing the cluster generation rate $\lambda_G$ and recombination rate $\lambda_R$.
The antenna patterns in \eqref{CIR_N} can be replaced with those of high-directional antennas, which are often used in mmWave-THz communications.

\section{Statistical Properties}
\subsection{Local STF-CF}
The local STF-CF between $H_{qp}(t,f)$ and $H_{\tilde q\tilde p}(t-\Delta t,f-\Delta f)$ is defined as
\begin{equation}\label{equ_STFCF}
R_{qp,\tilde q\tilde p}(t,f;\Delta r,\Delta t, \Delta f)=\mathbb E\{H_{qp}(t,f)H^*_{\tilde q\tilde p}(t-\Delta t,f-\Delta f)\}.
\end{equation}
By substituting (\ref{equ_TF}) into (\ref{equ_STFCF}), the STF-CF is further written as
\begin{align}\label{equ_STFCF2}
R_{qp,\tilde q\tilde p}&(t,f;\Delta r,\Delta t, \Delta f)= \frac{K_R}{K_R+1}R^L_{qp,\tilde q\tilde p}(t,f;\Delta r,\Delta t, \Delta f)\nonumber\\
&+\frac{1}{K_R+1}\sum_{n=1}^{N_{qp}(t)}R^N_{qp,\tilde q\tilde p,n}(t,f;\Delta r,\Delta t, \Delta f)
\end{align}
where the LoS and NLoS components of the STF-CF can be obtained as
\begin{align}\label{equ_STFCFLoS}
R^{L}_{qp,q\tilde p}(t,f;\Delta r,\Delta t, \Delta f)=[P^{L}_{qp}(t)P^ L_{\tilde q\tilde p}(t-\Delta t)]^{\frac{1}{2}}\nonumber \\ \times e^{j\frac{2\pi(f_c-f)}{\lambda f_c}[d^ L_{qp}(t)-d^ L_{\tilde q\tilde p}(t-\Delta t)]-j\frac{2\pi\Delta f}{\lambda f_c}d^ L_{\tilde q\tilde p}(t-\Delta t)}
\end{align}
\begin{align} \label{equ_STFCFNLoS}
&R^{ N}_{qp,q\tilde p,n}(t,f;\Delta r,\Delta t, \Delta f)=P_\text{remain}(\Delta t,\Delta r)
\cdot \mathbb E\{\sum_{m_n=1}^{M_n}a_{m_n} \nonumber \\ & e^{j\frac{2\pi(f_c-f)}{\lambda f_c}[d_{qp,m_n}(t)-d_{\tilde q\tilde p,m_n}(t-\Delta t)]-j\frac{2\pi\Delta f}{\lambda f_c}d_{\tilde q\tilde p,m_n}(t-\Delta t)}\}
\end{align}
where $\Delta r=\{\Delta r^T,\Delta r^R\}$, $\Delta r^T = \delta_{\tilde p} -\delta_p$, $\Delta r^R = \delta_{\tilde q} -\delta_ q$,  $a_{m_n}=[P_{qp,m_n}(t)P_{\tilde q\tilde p,m_n}(t-\Delta t)]^{\frac{1}{2}}[\frac{f(f-\Delta f)}{f_c^{2}}]^{\gamma_{m_n}}$, $P_\text{remain}(\Delta t,\Delta r)$ is the joint probability of a cluster survives from $t-\Delta t$ to $t$ on time axis and from $A^T_p$ to $A^T_{\tilde p}$ and from $A^R_q$ to $A^R_{\tilde q}$ on array axes.

For the stationary case, i.e., the model is stationary over $r$, $t$, and $f$.
We further assume that the delays within a cluster are irresolvable.
By setting $P_\text{remain}(\Delta t,\Delta r)=1$, $\gamma_{m_n}=0$, $a_{m_n}=1/M_n$ and removing the SWF and non-WSS terms in (\ref{TraDis1}), the NLoS components of STF-CF reduces to
\begin{align} \label{equ_STFCFNLoS2}
&R^ N_{qp,\tilde q\tilde p,n}(t,f;\Delta r,\Delta t,\Delta f)
= \mathbb E \{\sum_{m_n=1}^{M_n} \frac{1}{M_n} \nonumber \\
& \cdot  e^ {j\frac{2\pi(f_c-f)}{\lambda f_c} [\cos\vartheta^R\Delta r^R+\cos\vartheta^T\Delta r^T-(\cos\omega^Tv^T+\cos\omega^Rv^R)\Delta t]} \nonumber \\
&\cdot e^{- j\frac{2\pi\Delta f}{\lambda f_c}d_{\tilde q\tilde p,m_n}(t-\Delta t)}\}.
\end{align}
%
%
%
Note that the CF in this case is still STF-dependent.
By imposing $\Delta f=0$, i.e., removing the frequency selectivity, the CF becomes WSS in the space and time domains, i.e., only relies on $\Delta r$ and $\Delta t$.
Similarly, the CF is WSS in the frequency domain when the time selectivity and space selectivity are ignored, i.e., setting $\Delta t=0$, $\Delta r=0$.
\subsection{Local Spatial-Doppler PSD}
The local spatial-Doppler PSD can be obtained as the Fourier transform of spatial CF $R_{q,p\tilde p,n}(t,f;\Delta r,0,0)$ w.r.t. $\Delta r$. For the Tx side, the local spatial-Doppler PSD is obtained as
\begin{equation}\label{equ_SDPSD}
G_{q,p\tilde p,n}(t,f;\varpi )\!=\!\!\int\!\! R_{q,p\tilde p,n}(t,f;\Delta r^{T},0,0)e^{-j\frac{2\pi}{\lambda}\varpi \Delta r^T}\mathrm d \Delta r^{T}.
\end{equation}
Note that $\Delta r^R=0$ indicates two links share the same receive antenna.
The local spatial-Doppler PSD in \eqref{equ_SDPSD} describes the distribution of average power on the spatial-Doppler frequency axis at antenna $A^T_p$, time $t$, and frequency $f$.

\subsection{Local Doppler Spread}
The instantaneous frequency provides a measure of the energy distribution of a signal over the frequency domain and is important for signal recognition, estimation, and modeling.
The instantaneous frequency, which is given by the instantaneous Doppler frequency, is estimated as $\frac{\mathrm d \varphi(t)}{2\pi\mathrm d t}$, where $\varphi(t)$ accounts for the phase change of the channel \cite{Boashash1992InstantDoppler}.
Based on \eqref{CIR_N}, the instantaneous Doppler frequency of the proposed model is given by $\nu_{qp,m_n}(t)=\frac{1}{\lambda}\frac{\mathrm d{[d_{p,m_n}^T(t)+d_{m_n,q}^R(t)]}}{\mathrm d t}$, and is further expressed as
\begin{align}\label{equ_Doppler2}
&\nu_{qp,m_n}(t)=-\frac{v^T}{\lambda}\cos(\omega^T_p)-\frac{v^R}{\lambda}\cos(\omega^R_q)\nonumber \\ &+ \frac{\sin^2(\omega^T_p)(v^T)^2 t}{\lambda(d^T_{m_n}-\cos(\vartheta^T)\delta_p)}+\frac{\sin^2(\omega^R_q)(v^R)^2t}{\lambda (d^R_{m_n}-\cos(\vartheta^R)\delta_q)}.
\end{align}
Note that the instantaneous Doppler frequency varies with time caused by the movements of the Tx, Rx, and scatterers.
The advantage of \eqref{equ_Doppler2} w.r.t. other channel models such as \cite{Yuan2015V2V} and \cite{Wu2017unified} lies in that the Doppler frequency can be written as the summation of two components.
The first two terms of \eqref{equ_Doppler2} are the conventional Doppler frequency expression in stationary case.
The last two terms of \eqref{equ_Doppler2} accounts for the time-variation of the Doppler frequency in the non-stationary case.
Finally, the local Doppler spread can be calculated as
\begin{equation}\label{equ_DopplerSpread}
B_{qp}^{(2)}(t)=\left( \mathbb E[\nu_{qp,m_n}(t)^2]-\mathbb E[\nu_{qp,m_n}(t)]^2\right )^{\frac{1}{2}}.
\end{equation}

\subsection{Array Coherence Distance}
As a counterpart of coherence time and coherence bandwidth, the array coherence distance is the minimum antenna element spacing during which the spatial CF equals to a given threshold $c_\text{thresh}$.
The transmit antenna array coherence distance can be expressed as \cite{Fleury2000SpatialDoppler}
\begin{equation}\label{equ_CohDis}
I_p(c_\text{thresh})=\text{min}\{\left | \Delta r^T \right |:\left | R_{q,p\tilde p,n}(t,f;\Delta r^T,0,0)\right |=c_\text{thresh} \}
\end{equation}
where $c_\text{thresh} \in [0,1]$.
The receive antenna array coherence distance can be calculated similarly.
In (\ref{equ_CohDis}), small values of $c_\text{thresh}$ results in the minimum distance between two antenna elements over which the channels can be considered as independent.
However, larger values of $c_\text{thresh}$ provide the maximum antenna spacing within which the channels do not change significantly.

\section{Results and Analysis}
In this section, results of important statistics of the B5GCM are presented.
Some of the statistics including spatial CF, cluster VR length, cluster power variation, Doppler spread, and delay spread are compared with the corresponding channel measurement data.
In the simulation, the parameters such as carrier frequency, antenna height, Tx-Rx separation, and velocity of the Tx/Rx are set according to the corresponding measurements.
Only a small number of parameters, such as $\lambda_R$, $D_c^A$, $\sigma_{n}$, $\sigma_{AS}$, $\sigma_{ES}$, and $\sigma_{DS}$,  which differentiate the proposed model as compared to conventional ones, are determined by fitting statistical properties to the channel measurement data.
The rest of parameters are randomly generated according to the 3GPP TR38.901 channel model.
Unless otherwise noted, the following parameters are used for simulation: $f_c=2.6$~GHz, $M_T=128$, $M_R=1$, $\beta_A^T = \pi/6$, $\beta_E^T=0$, $\mu=1$, $D=100$ m, $\lambda_R=6.79$/m, $\lambda_G=81.56$/m, $D_c^A=9.93$~m, $\sigma_n = 0.054$. In addition, and half-wave dipole antennas with vertical polarization are adopted in simulations.

\begin{figure}
\begin{minipage}[t]{1\linewidth}
\centering
\includegraphics[width=3.5in]{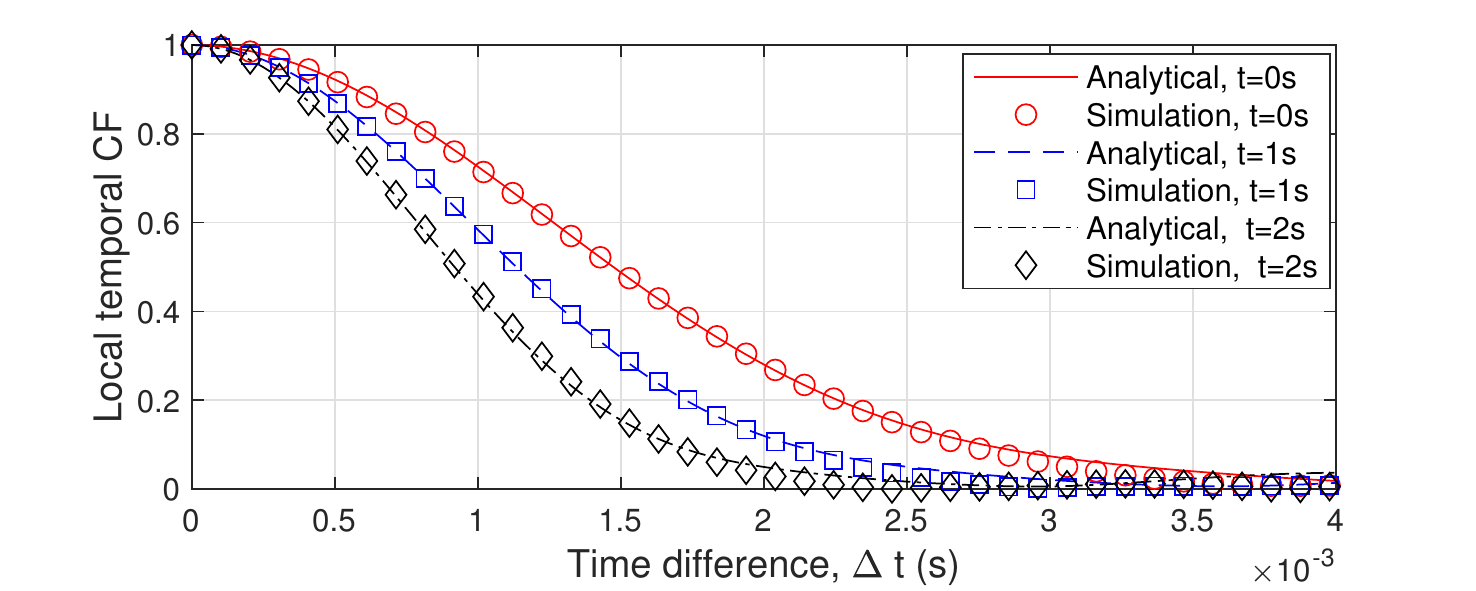} \vspace{-0.5cm}
\caption*{\small{(a) }}
\end{minipage} \\
\begin{minipage}[t]{1\linewidth}
\centering
\includegraphics[width=3.5in]{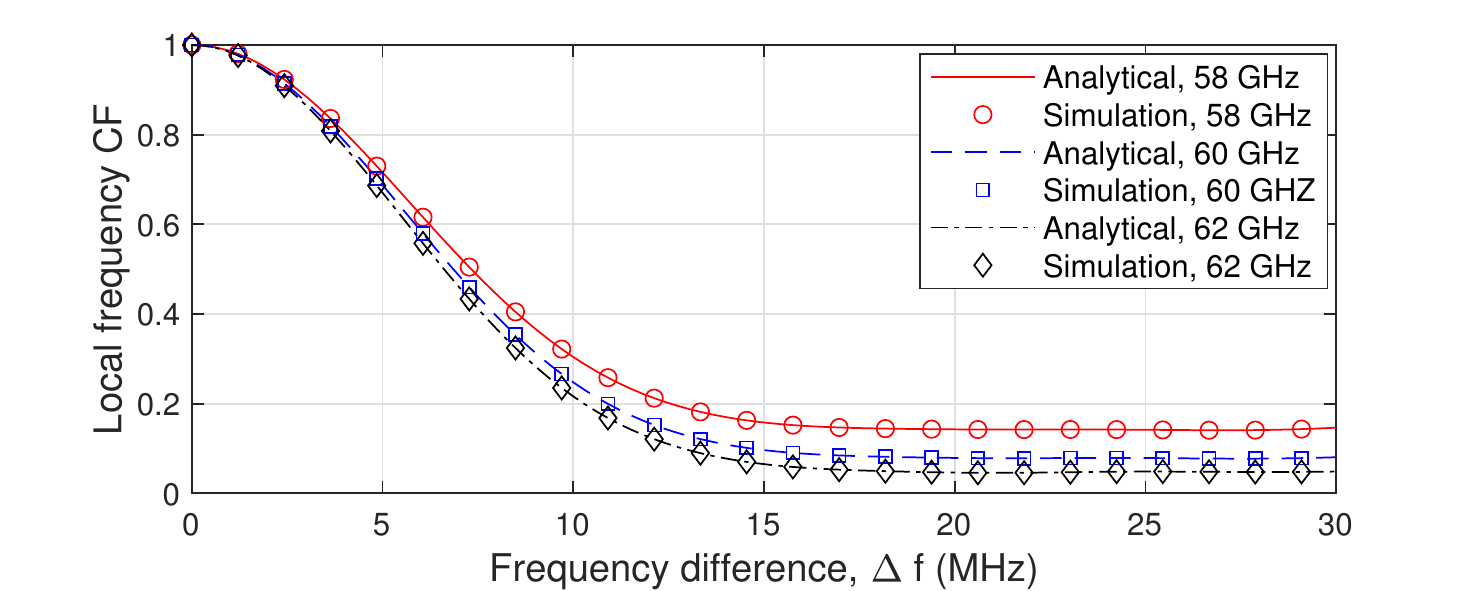} \vspace{-0.5cm}
\caption*{\small{(b)}}
\end{minipage}
\begin{minipage}[t]{1\linewidth}
\centering
\includegraphics[width=3.5in]{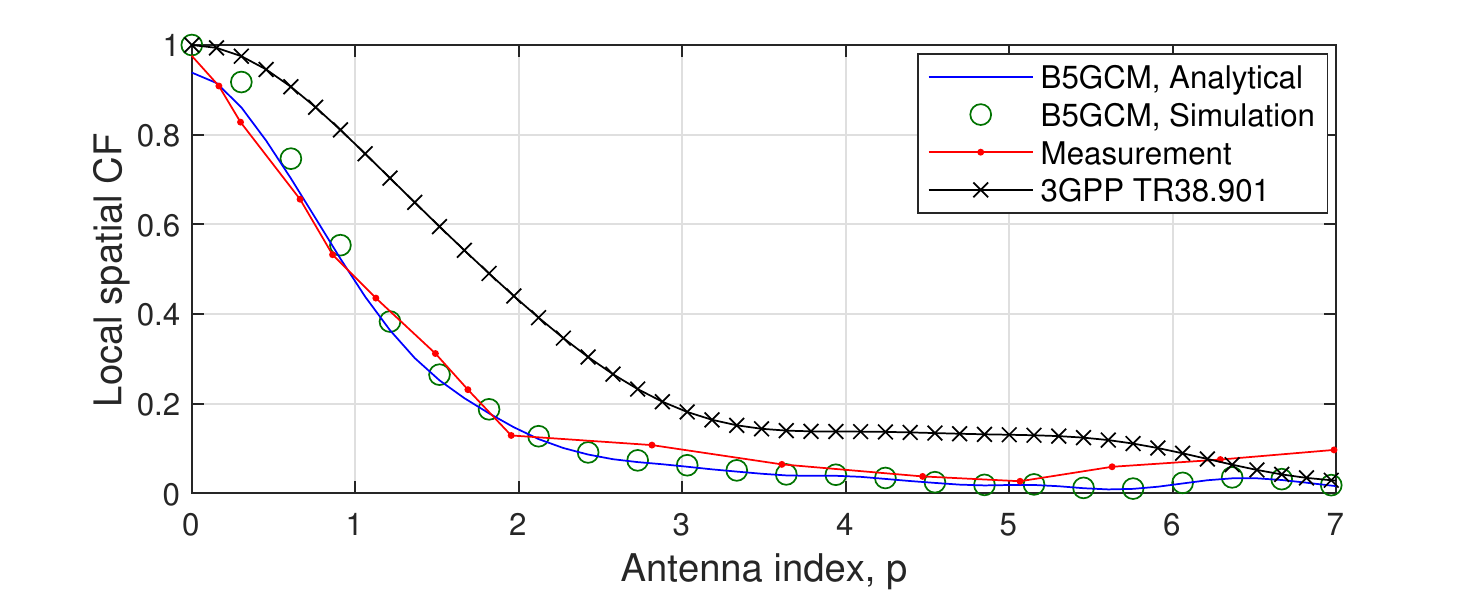} \vspace{-0.5cm}
\caption*{\small{(c)}}
\end{minipage}
\caption{Local temporal, frequency, and spatial CFs of the B5GCM, and measurement data in \cite{Payami2012MaMIMO_meas} ($v^{T(R)} = 10$~m/s $\sigma_{DS}=6.82$ m, $\sigma_{AS}=11.68$ m, $\sigma_{ES}=9.21$ m).}
\label{fig_CF}
\end{figure}

For the parameters listed above, the local temporal, frequency, and spatial CFs of the proposed model are shown in Fig.~\ref{fig_CF}.
Specifically, the local temporal CFs at 0~s, 1~s and 2~s are shown in Fig.~\ref{fig_CF}(a).
Note that the analytical results are generated by imposing $\Delta r=0$ and $\Delta f=0$ in \eqref{equ_STFCF}.
The simulation results are obtained based on two channel transfer functions separated by different time.
The time-variations of temporal CFs result from the motions of the Tx, Rx and the survival probability of the cluster, which make the model non-stationary in the time domain.
Fig. \ref{fig_CF}(b) presents the frequency CFs of the B5GCM.
The frequency CFs vary with frequency due to the frequency dependence of the path gains, indicating the frequency non-stationarity of the proposed model.
Fig.~\ref{fig_CF}(c) provides the comparison of the local spatial CFs of the B5GCM, 3GPP TR38.901 \cite{3GPP38901}, and the measurement data \cite{Payami2012MaMIMO_meas}.
Note that the space differences have been normalized w.r.t. antenna spacing.
The measurement was carried out at 2.6~GHz in a court yard scenario, where a 7.3 m 128-element virtual ULA is used.
The antenna forming the virtual ULA is spaced at half-wavelength and illustrates omnidirectional pattern in the azimuth plane.
The result shows that the spatial CFs of the B5GCM provide a better consistency with the measurement data than those of the 3GPP model.
This is because the 3GPP model neglected the effect of SWF. Besides, the non-stationarity over large antenna array was neglected.

\begin{figure}
\centering\includegraphics[width=3.5in]{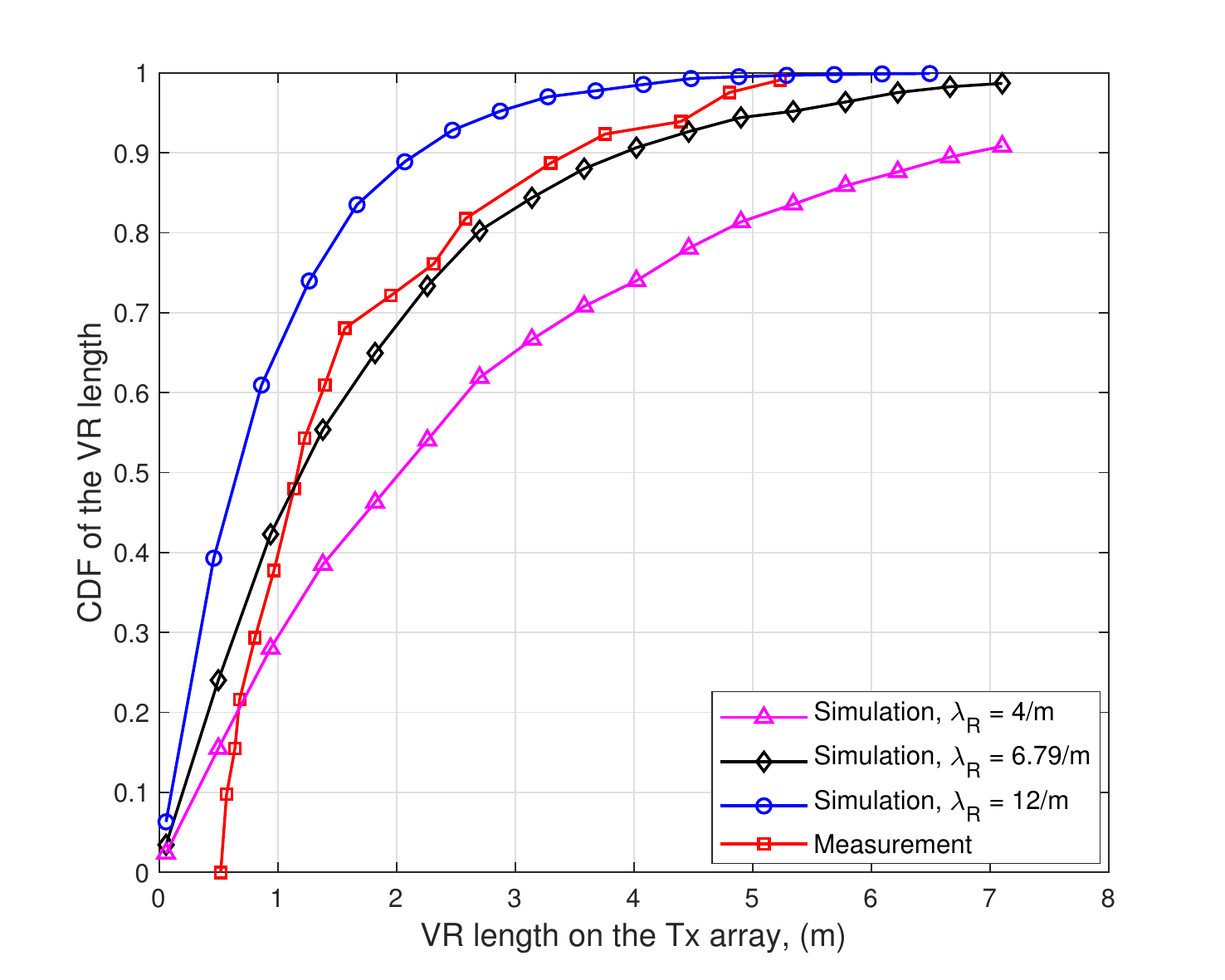}
\caption{CDF of the VR length of the proposed B5GCM and the measurement data in \cite{Gao2013MaMIMO_meas} ($\lambda_R=6.79$/m, $\lambda_G=81.56$/m, $D_c^A=9.93$~m, $\sigma_n=0.054$).}
\label{fig_VR}
\end{figure}
\begin{figure} [!htb]
\centering\includegraphics[width=3.5in]{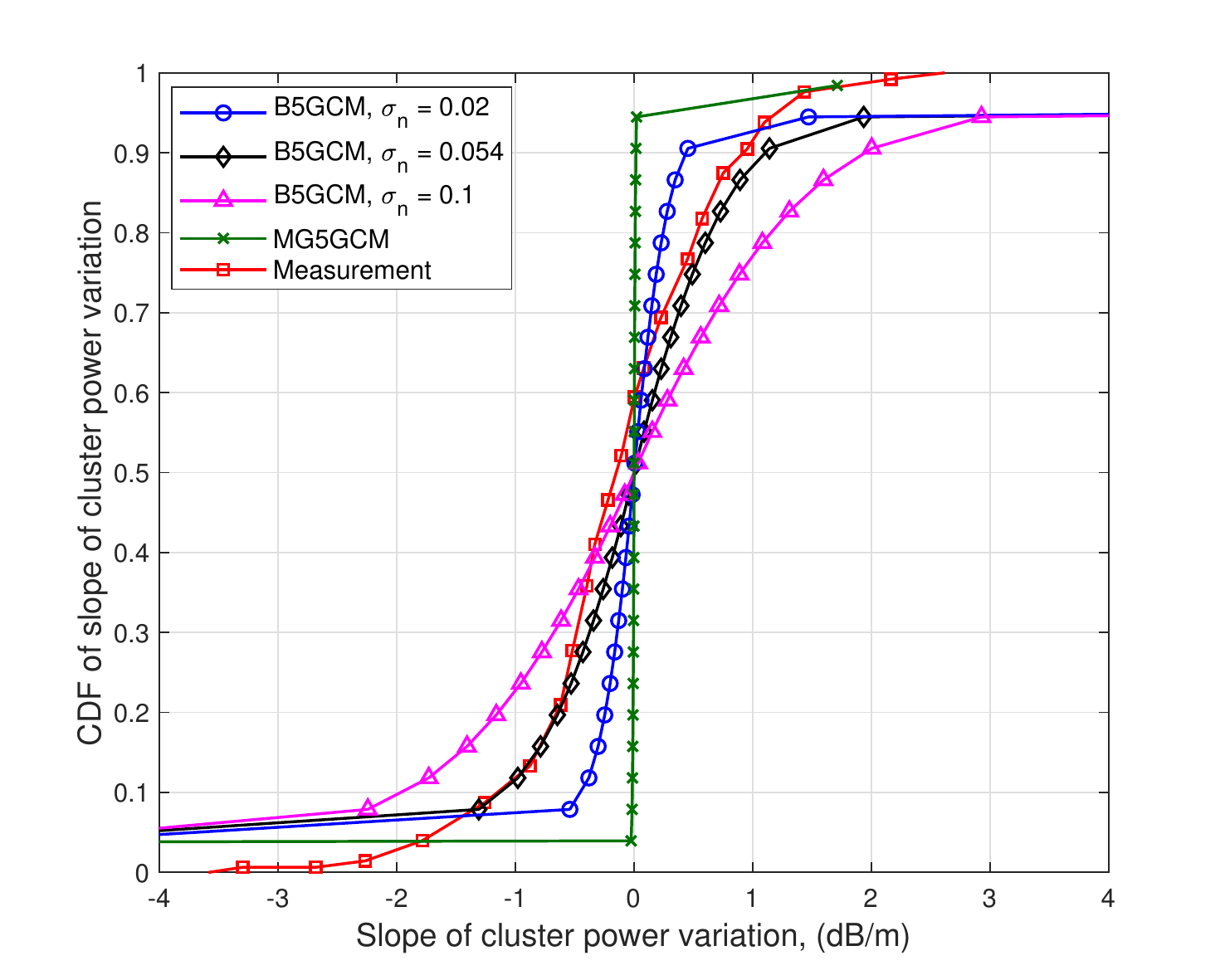}
\caption{CDF of the slopes of cluster power variations of the proposed B5GCM, MG5GCM in \cite{Wu2017unified}, and the measurement data in \cite{Gao2013MaMIMO_meas} ($\lambda_R=6.79$/m, $\lambda_G=81.56$/m, $D_c^A=9.93$~m, $\sigma_n=0.054$).}
\label{fig_PowerVariation}
\end{figure}

The simulated cumulative distribution functions (CDFs) of VR length and slope of cluster power variation on the array are presented in Figs. \ref{fig_VR} and \ref{fig_PowerVariation}, respectively.
The measurement used for comparison was carried out at 2.6 GHz in a campus scenario, where an omnidirectional antenna moves along a rail with a half-wavelength spacing, constituting a 128-element virtual ULA \cite{Gao2013MaMIMO_meas}.
The model parameters were chosen by minimizing the error norm $\varepsilon= \sum_{m=1}^{2}w_m\mathbb E\{| \hat F_m-F_m(\mathcal P)|^2\}$, where $\hat F_m$ and $F_m$ are the measured and derived statistics, respectively, $w_m$ is the weight of the $m$th error norm and satisfying $w_1+w_2=1$, $\mathcal P=\{\lambda_R,D_c^A,\sigma _n\}$ is parameter set to be jointly optimized.
Note that we impose $\lambda_G/\lambda_R=12$ to ensure a constant cluster number along the array.
It is found that $\lambda_R=6.79$/m, $D_c^A=9.93$~m, and $\sigma_n=0.054$ can be chosen as a good match.
The results show that increasing the cluster recombination rate leads to a shorter VR, which indicates a larger spatial non-stationarity.
Furthermore, results in Fig. \ref{fig_PowerVariation} suggest that large values of $\sigma_n$ can increase the cluster power variation over the array.
However, the channel model in \cite{Wu2017unified} assumed that the cluster powers are constant along the array, which may underestimate the spatial non-stationarity of massive MIMO channels.

\begin{figure}
\centering\includegraphics[width=3.5in]{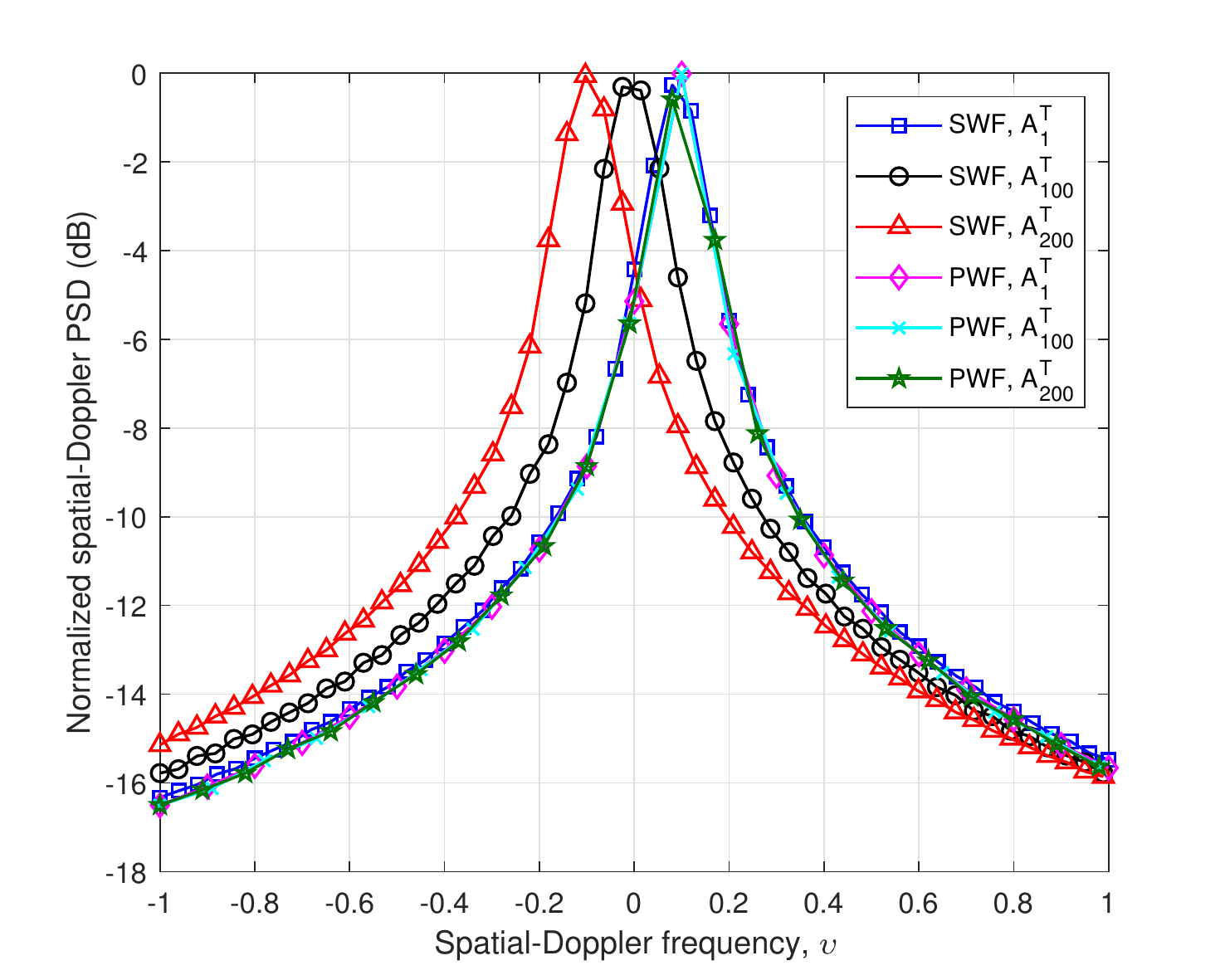}
\caption{The simulated normalized spatial-Doppler PSDs of the B5GCM at $A^T_1$, $A^T_{100}$, and $A^T_{200}$ using SWF and PWF assumptions ($\phi_A^T = 2\pi/3$, $\phi_E^T=\pi/9$, $\lambda_R=6.79$/m, $\lambda_G=81.56$/m, $D_c^A=9.93$~m, $\sigma_n=0.054$).}
\label{fig_spatialDoppler}
\end{figure}

\begin{figure}
\centering\includegraphics[width=3.5in]{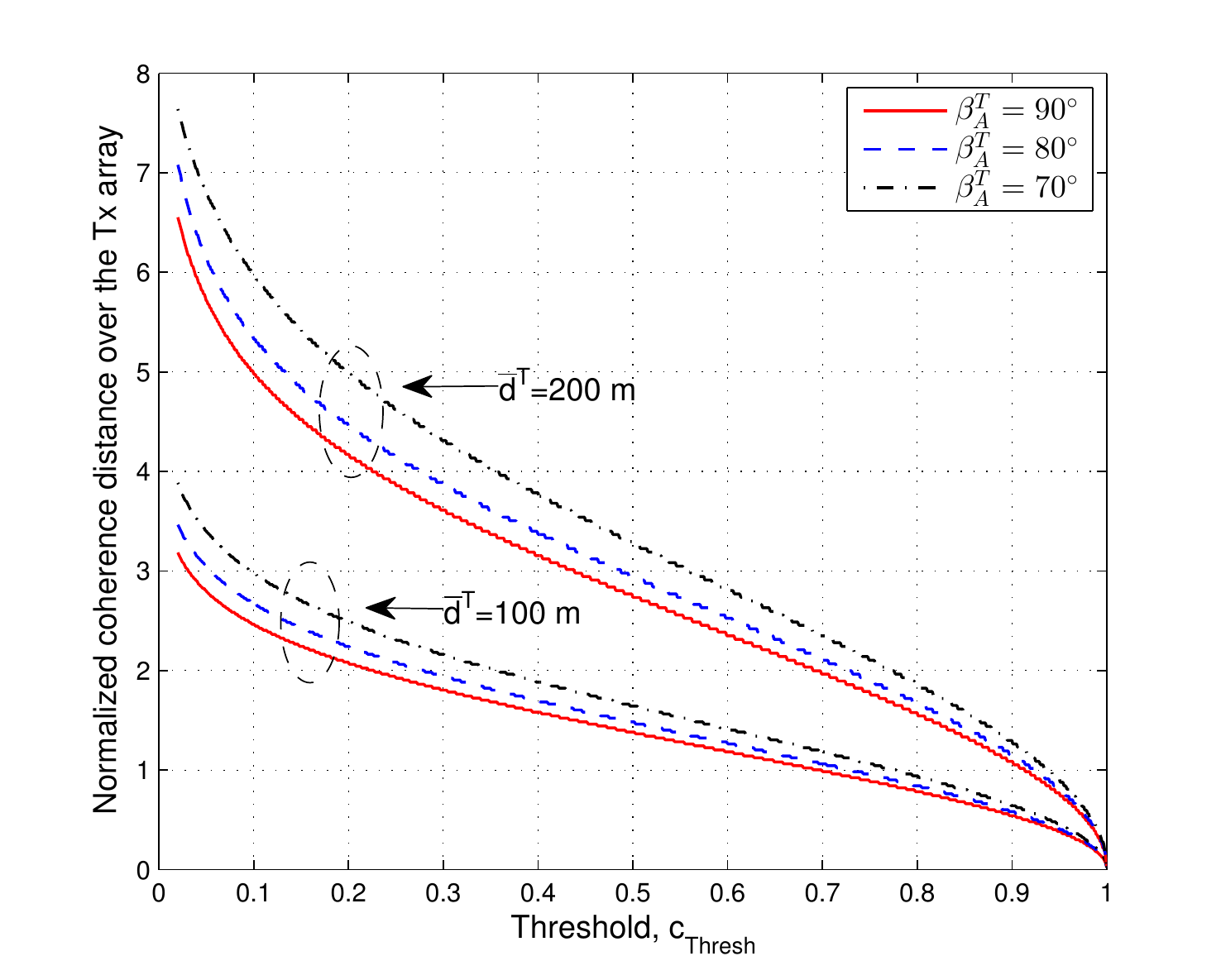}
\caption{Simulated coherence distances of the B5GCM over array ($\phi_A^T = \pi/10$, $\phi_E^T=\pi/9$, $\lambda_R=6.79$/m, $\lambda_G=81.56$/m, $D_c^A=9.93$~m, $\sigma_n=0.054$).}
\label{fig_CoherDis}
\end{figure}

Fig.~\ref{fig_spatialDoppler} shows the simulated normalized spatial-Doppler PSDs, which are obtained according to \eqref{equ_SDPSD}.
For the SWF case, the variations of spatial-Doppler PSDs along the transmit array are caused by the large size of the transmit antenna aperture.
However, for the PWF case, the values of spatial-Doppler PSDs are constant over the array, which may result in inaccurate performance estimations of massive MIMO systems.
Besides, Fig. \ref{fig_CoherDis} shows the simulated array coherence distance based on \eqref{equ_CohDis}.
Note that the coherence distances have been normalized w.r.t. antenna spacing.
The results indicate that the spatial non-stationarity, which is caused by SWF and cluster array evolution, is affected by both the array orientation and the Tx/Rx-cluster separation.
The channel has a shorter array coherence distance when distance from the Tx/Rx to the cluster decreases.
Furthermore, increasing the angles between array orientation and rays can lead to a larger array coherence distance.

\begin{figure} \label{fig_DopplerSpread}
\centering\includegraphics[width=3.5in]{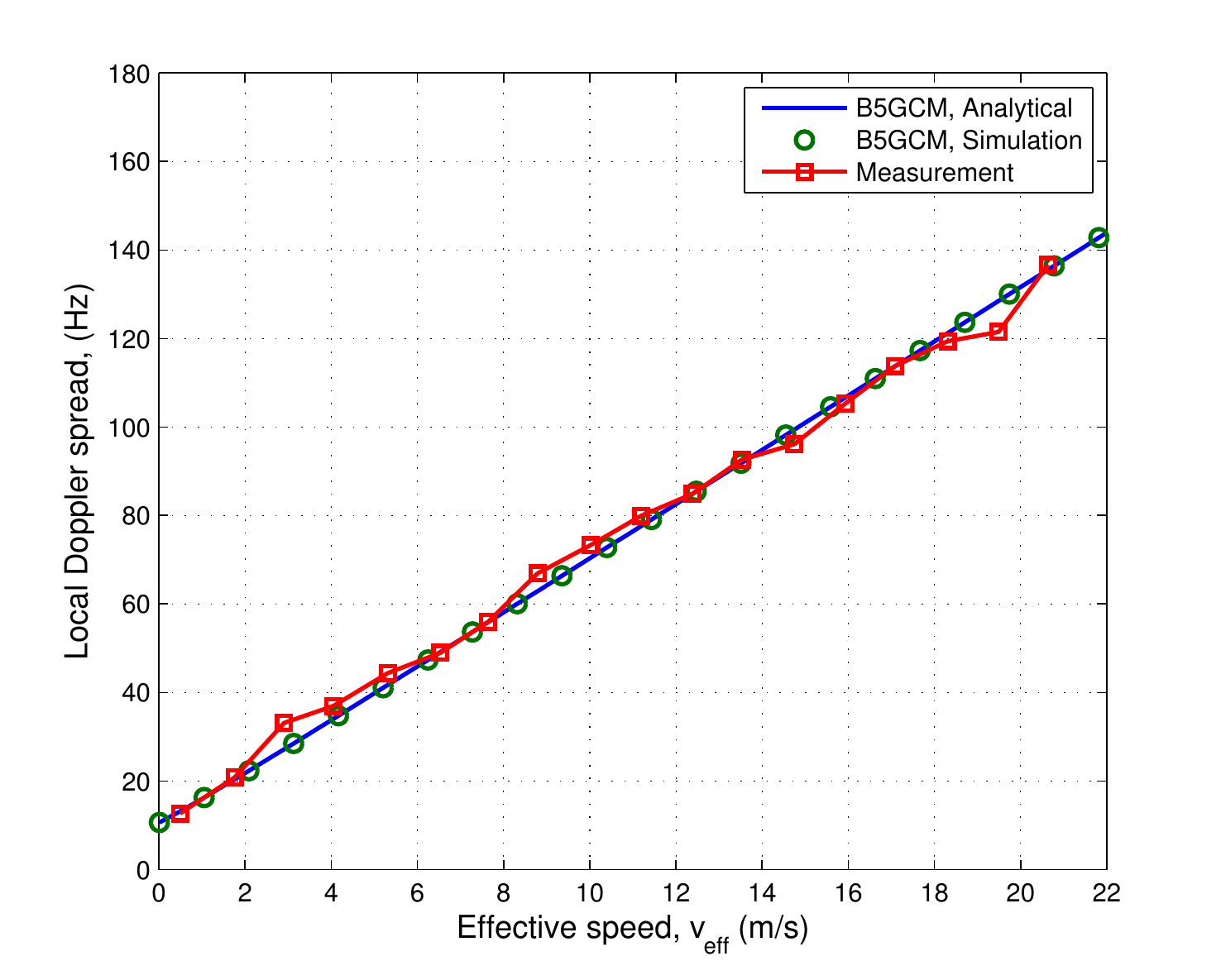}
\caption{Doppler spread of the B5GCM versus effective speed and the measurement data in \cite{Cheng2008V2V_meas} ($f_c=5.9$~GHz, $\sigma_{DS}=81.04$ m, $\sigma_{AS}=88.92$ m, $\sigma_{ES}=72.03$ m,  $\alpha^{T}=0$, $\alpha^{R}=\pi$, $v^{A_n}=0$~m/s, $v^{Z_n}=2$~m/s, $\alpha^{Z_n}=0$).}
\label{fig_DopplerSpread}
\end{figure}

Fig. \ref{fig_DopplerSpread} compares the Doppler spread of the B5GCM with the measurement data \cite{Cheng2008V2V_meas}.
The Doppler spread is obtained according to \eqref{equ_DopplerSpread}.
The channel measurement was conducted at 5.9~GHz in highway, rural, and suburban environments.
The $x$-axis of this figure is the effective speed, which is defined as $v_\text{eff}=[{(v^T)}^2+{(v^R)}^2]^{\frac{1}{2}}$.
A good consistency among the simulated, analytical results, and the corresponding measurement data can be observed.
Noting that the Doppler spread illustrates a nonzero value when $v_\text{eff}=0$.
It stems from the extra Doppler shifts due to the motion of scatterers, and cannot be obtained by the models assuming static clusters \cite{WINNERPlus,3GPP3D,3GPP38901}.

\begin{figure}
\centering\includegraphics[width=3.5in]{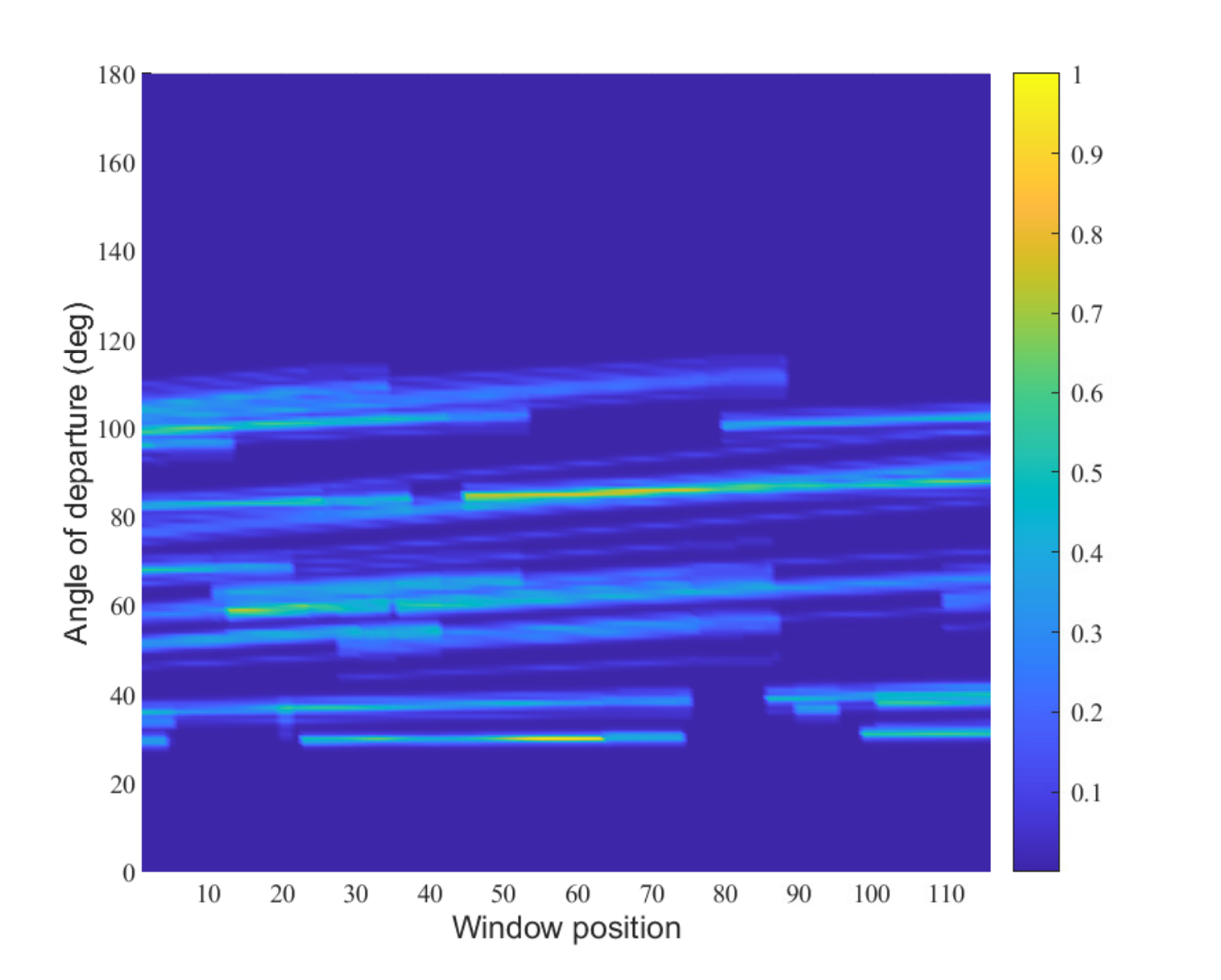}
\caption{A snapshot of the APS of AoD of the B5GCM ($\lambda_R=6.79$/m, $\lambda_G=81.56$/m, $D_c^A=9.93$~m, $\sigma_n=0.054$).}
\label{fig_APS}
\end{figure}

The simulated angle power spectrum (APS) of AoD of the B5GCM is shown in Fig. \ref{fig_APS}.
The result is obtained using the multiple signal classification (MUSIC) algorithm \cite{Schmidt1986MUSIC}.
A sliding window consisting of 12 consecutive antennas is shifted along the array in order to capture the channel non-stationaries in the space domain.
Besides, the birth-death process of clusters along the transmit array can be seen.
Some clusters with strong powers are observable along the whole array.
Other weak power clusters only appear to part of the array.
The power of clusters vary smoothly over the array can be observed due to the spatial lognormal process.
Moreover, the angles of rays experience linear drifts along the array caused by the nearfield effects, which has been validated by several channel measurement campaigns \cite{Payami2012MaMIMO_meas,Li2019MaMIMO}.

\begin{figure}[!htb]
\centering\includegraphics[width=3.5in]{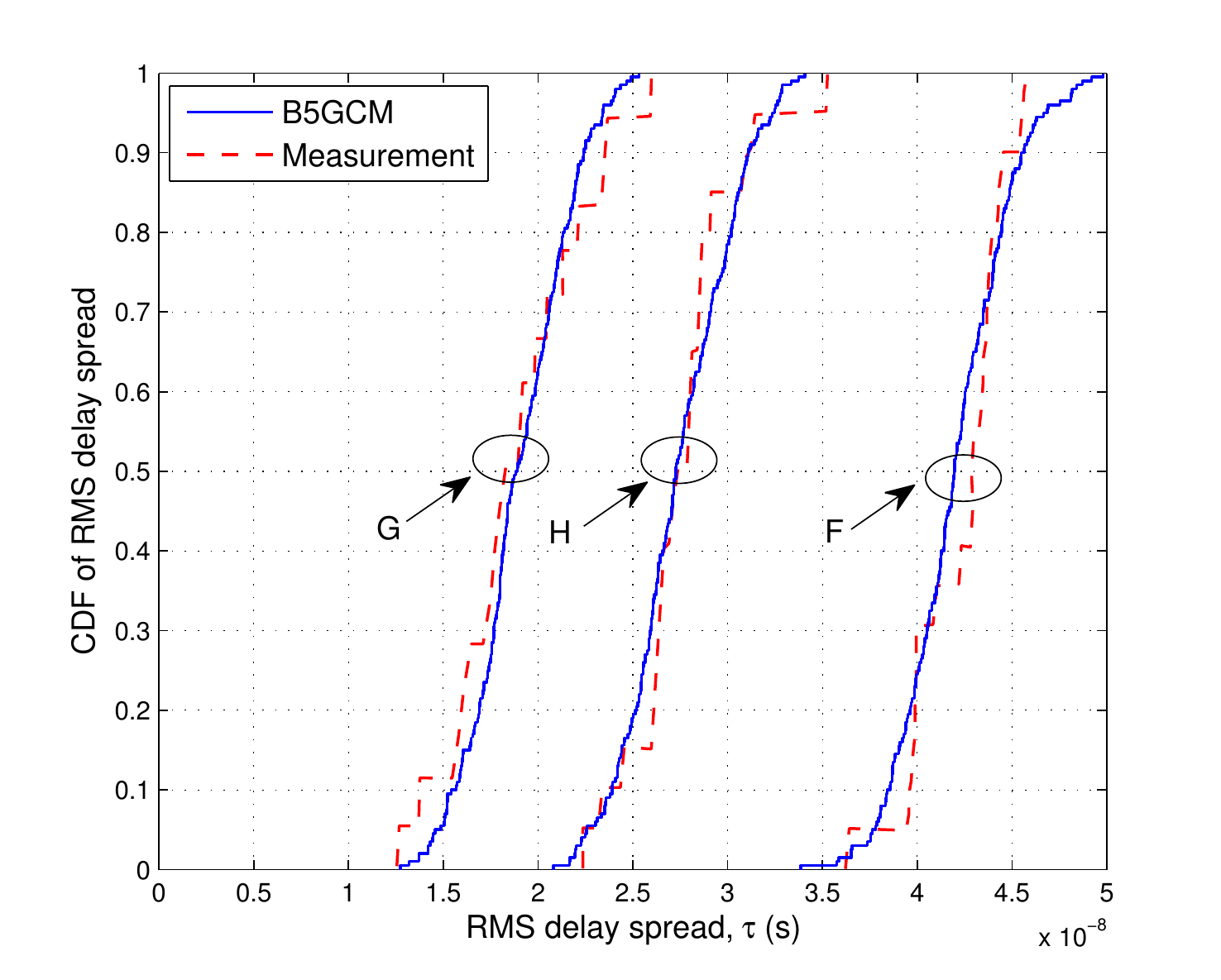}
\caption{RMS delay spread statistics of the proposed B5GCM and measurement data in \cite{Smulders1995} ($f_c=58$ GHz, $D=3$~m, $\overline d^T \sim \mathcal N(3,0.5)$ m,  Room G: $\sigma_{DS}=1.1$ m, $\sigma_{AS}=1.4$ m, $\sigma_{ES}=1.4$ m, Room H: $\sigma_{DS}=2.3$ m, $\sigma_{AS}=1.8$ m, $\sigma_{ES}=1.4$ m, Room F: $\sigma_{DS}=3.8$~m, $\sigma_{AS}=2.1$ m, $\sigma_{ES}=1.1$ m. ) }
\label{fig_DS}
\end{figure}
The CDF of the RMS delay spreads of the B5GCM and the measurement data in \cite{Smulders1995} are compared in Fig. \ref{fig_DS}.
The measurements were conducted at 58~GHz in three indoor scenarios, i.e., Lecture room (Room G), Laboratory room (Room H), and Lecture room (Room F).
Both the Tx and Rx antennas are equipped with motionless bicone antennas.
The different RMS delay spreads for the three propagation environments are caused by different distributions of scatterers within cluster.
Good agreements between the results of the B5GCM and measurement data show the usefulness of the proposed model.

\section{Conclusions}
This paper has proposed a novel 3D STF non-stationary GBSM for 5G and B5G wireless communication systems.
The proposed model is applicable to various communication scenarios, e.g., massive MIMO, HST, V2V, and mmWave-THz communication scenarios.
Important (B)5G channel characteristics have been integrated, including SWF, cluster power variation over array, Doppler shifts caused by motion of scatterers, time-variant velocity and trajectory, and spatial consistency.
Note that the above-mentioned channel characteristics have not been fully considered in the current 5G channel models, e.g., MG5GCM \cite{Wu2017unified}, 3GPP TR38.901 \cite{3GPP38901}, and IMT-2020 channel models \cite{IMT-2020}.
Furthermore, this paper has presented a general modeling framework.
The model can reduce to a variety of simplified channel models according to channel properties of specific scenarios, or be applied to new communication scenarios by setting appropriate model parameters.
Key statistics of the proposed model have been derived, some of which have been validated by measurement data, illustrating the generalization and usefulness of the proposed model.

\ifCLASSOPTIONcaptionsoff
  \newpage
\fi


\iftrue

\fi

\begin{IEEEbiography}[{\includegraphics[width=1.1in,height=1.3in,clip,keepaspectratio] {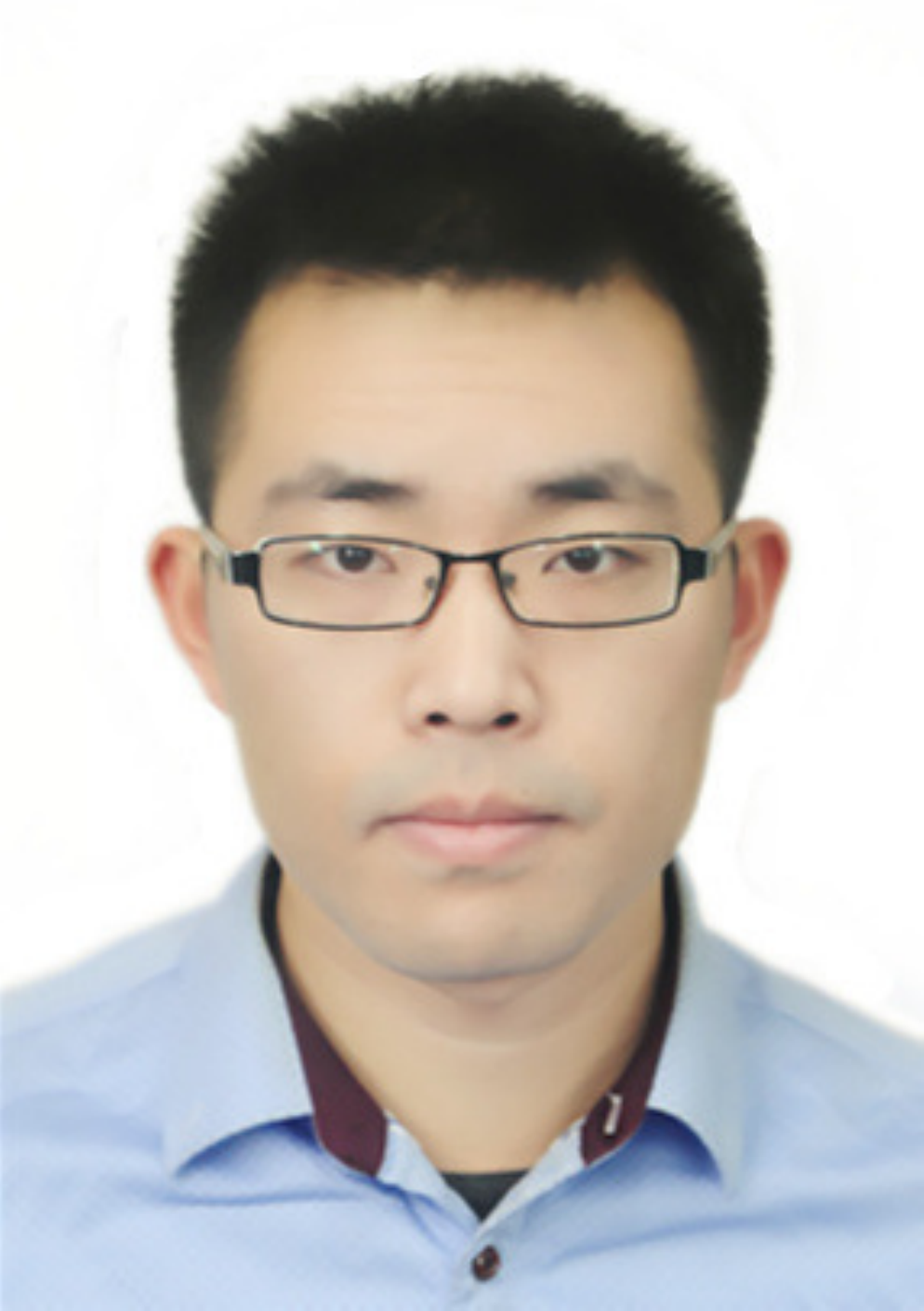}}]
{Ji Bian (M'20)} received the B.Sc. degree in electronic information science and technology from Shandong Normal University, Jinan, China, in 2010, the M.Sc. degree in signal and information processing from Nanjing University of Posts and Telecommunications, Nanjing, China, in 2013, and the Ph.D. degree in information and communication engineering from Shandong University, Jinan, China, in 2019.
From 2017 to 2018, he was a visiting scholar with the School of Engineering and Physical Sciences, Heriot-Watt University, Edinburgh, U.K.
He is currently a lecturer with the School of Information Science and Engineering, Shandong Normal University, Jinan, China. His research interests include 6G channel modeling and wireless big data.
\end{IEEEbiography}

\begin{IEEEbiography}[{\includegraphics[width=1.1in,height=1.3in,clip,keepaspectratio] {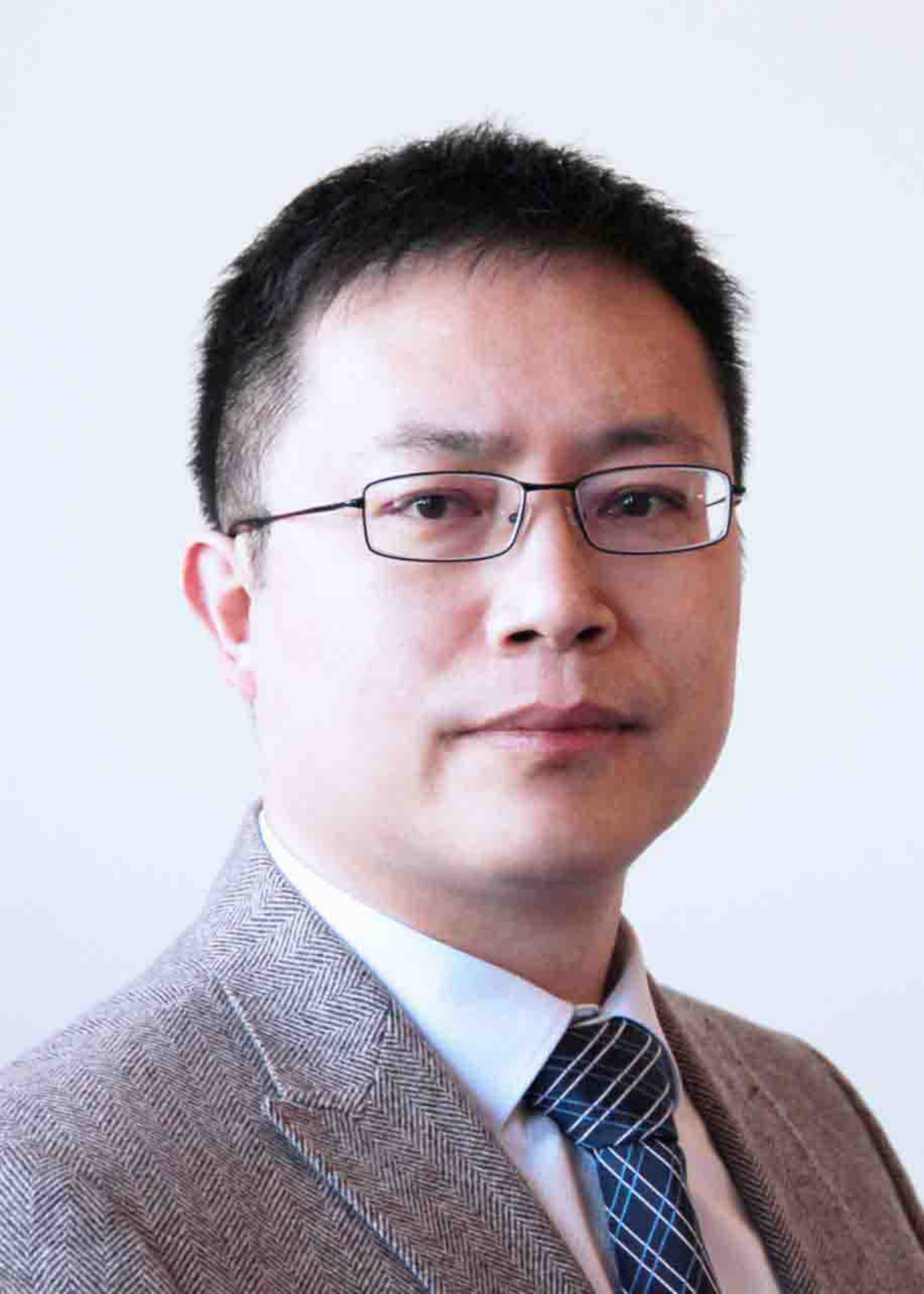}}]
{Cheng-Xiang Wang (S'01-M'05-SM'08-F'17)} received the BSc and MEng degrees in Communication and Information Systems from Shandong University, China, in 1997 and 2000, respectively, and the PhD degree in Wireless Communications from Aalborg University, Denmark, in 2004.

He was a Research Assistant with the Hamburg University of Technology, Hamburg, Germany, from 2000 to 2001, a Visiting Researcher with Siemens AG Mobile Phones, Munich, Germany, in 2004, and a Research Fellow with the University of Agder, Grimstad, Norway, from 2001 to 2005.
He has been with Heriot-Watt University, Edinburgh, U.K., since 2005, where he was promoted to a Professor in 2011. In 2018, he joined Southeast University, China, as a Professor.
He is also a part-time professor with the Purple Mountain Laboratories, Nanjing, China.
He has authored four books, three book chapters, and more than 410 papers in refereed journals and conference proceedings, including 24 Highly Cited Papers. He has also delivered 22 Invited Keynote Speeches/Talks and 7 Tutorials in international conferences.
His current research interests include wireless channel measurements and modeling, 6G wireless communication networks, and applying artificial intelligence to wireless communication networks.

Prof. Wang is a Member of the Academia Europaea (The Academy of Europe), a fellow of the IET, an IEEE Communications Society Distinguished Lecturer in 2019 and 2020, and a Highly-Cited Researcher recognized by Clarivate Analytics, in 2017-2020. He is currently an Executive Editorial Committee member for the IEEE TRANSACTIONS ON WIRELESS COMMUNICATIONS.
He has served as an Editor for nine international journals, including the IEEE TRANSACTIONS ON WIRELESS COMMUNICATIONS from 2007 to 2009, the IEEE TRANSACTIONS ON VEHICULAR TECHNOLOGY from 2011 to 2017, and the IEEE TRANSACTIONS ON COMMUNICATIONS from 2015 to 2017.
He was a Guest Editor for the IEEE JOURNAL ON SELECTED AREAS IN COMMUNICATIONS, Special Issue on Vehicular Communications and Networks (Lead Guest Editor), Special Issue on Spectrum and Energy Efficient Design of Wireless Communication Networks, and Special Issue on Airborne Communication Networks. He was also a Guest Editor for the IEEE TRANSACTIONS ON BIG DATA, Special Issue on Wireless Big Data, and is a Guest Editor for the IEEE TRANSACTIONS ON COGNITIVE COMMUNICATIONS AND NETWORKING, Special Issue on Intelligent Resource Management for 5G and Beyond. He has served as a TPC Member, TPC Chair, and General Chair for over 80 international conferences. He received twelve Best Paper Awards from IEEE GLOBECOM 2010, IEEE ICCT 2011, ITST 2012, IEEE VTC 2013-Spring, IWCMC 2015, IWCMC 2016, IEEE/CIC ICCC 2016, WPMC 2016, WOCC 2019, IWCMC 2020, and WCSP 2020.
\end{IEEEbiography}

\begin{IEEEbiography}[{\includegraphics[width=1.1in,height=1.3in,clip,keepaspectratio]{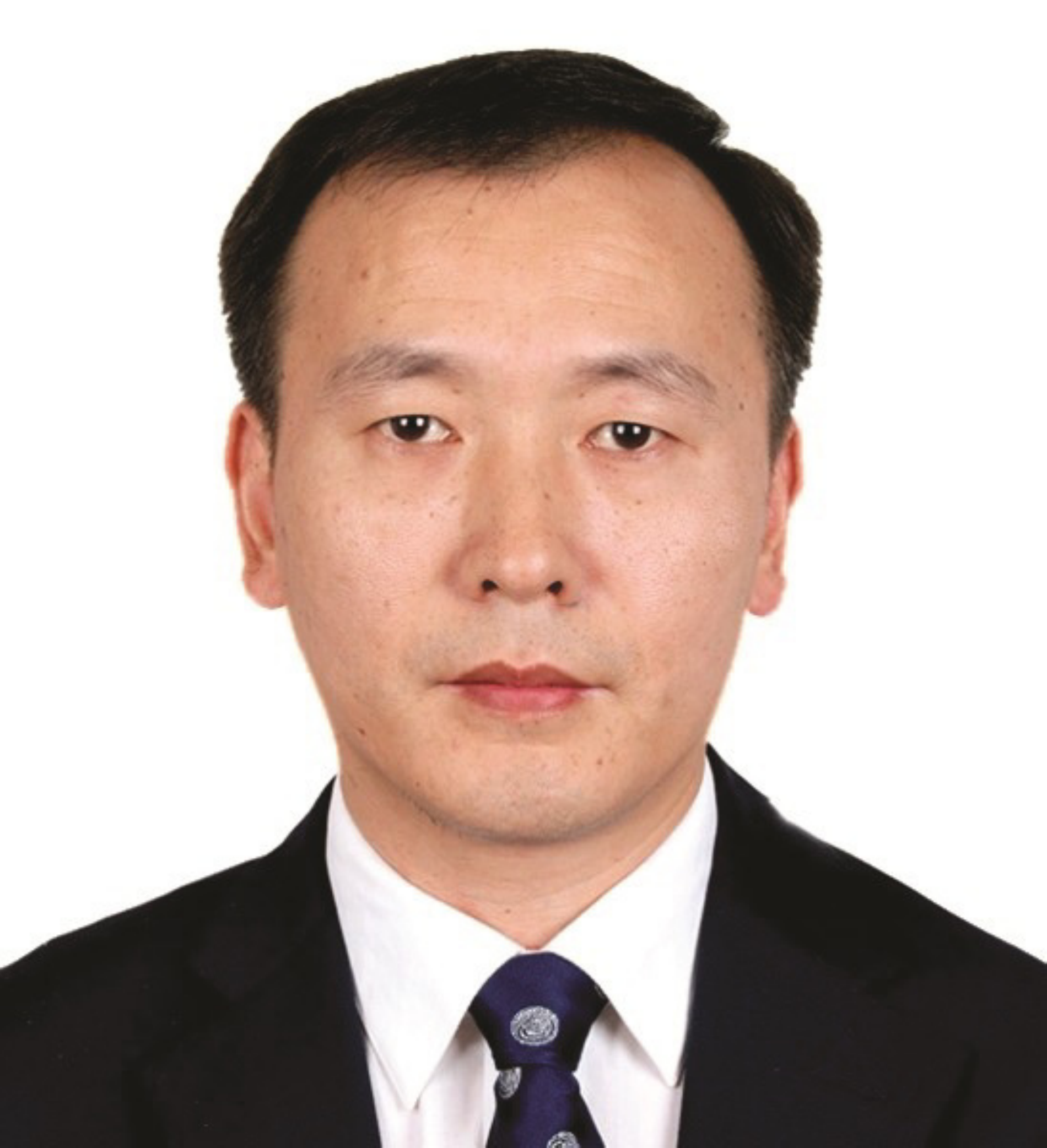}}]
{Xiqi Gao (S'92-AM'96-M'02-SM'07-F'15)} received the Ph.D. degree in electrical engineering from Southeast University, Nanjing, China, in 1997.
He joined the Department of Radio Engineering, Southeast University, in April 1992. Since May 2001, he has been a professor of information systems and communications. From September 1999 to August 2000, he was a visiting scholar at Massachusetts Institute of Technology, Cambridge, and Boston University, Boston, MA. From August 2007 to July 2008, he visited the Darmstadt University of Technology, Darmstadt, Germany, as a Humboldt scholar. His current research interests include broadband multi-carrier communications, MIMO wireless communications, channel estimation and turbo equalization, and
multi-rate signal processing for wireless communications. From 2007 to 2012, he served as an Editor for the IEEE TRANSACTIONS ON WIRELESS COMMUNICATIONS. From 2009 to 2013, he served as an Associate Editor for the IEEE TRANSACTIONS ON SIGNAL PROCESSING. From 2015 to 2017, he served as an Editor for the IEEE TRANSACTIONS ON COMMUNICATIONS.

Dr. Gao received the Science and Technology Awards of the State Education Ministry of China in 1998, 2006 and 2009, the National Technological Invention Award of China in 2011, and the 2011 IEEE Communications Society Stephen O. Rice Prize Paper Award in the field of communications
theory.
\end{IEEEbiography}

\begin{IEEEbiography}[{\includegraphics[width=1.1in,height=1.3in,clip,keepaspectratio]{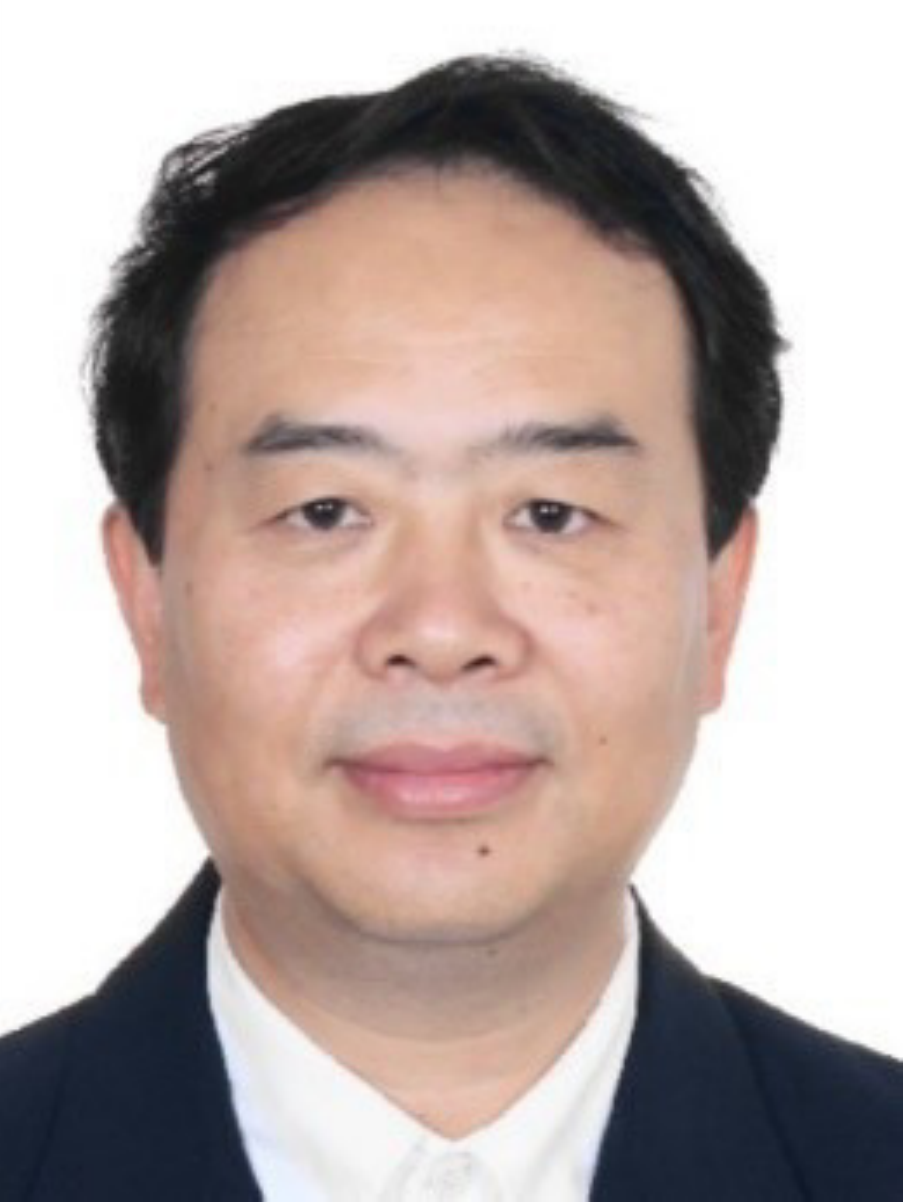}}]
{Xiaohu You (SM'11-F'12)}
has been working with National Mobile Communications Research Laboratory at Southeast University, where now he holds the rank of director and professor. He has contributed over 300 IEEE journal papers and 3 books in the areas of signal processing and wireless communications. From 1999 to 2002, he was the Principal Expert of the C3G Project. From 2001--2006, he was the Principal Expert of the China National 863 Beyond 3G FuTURE Project. Since 2013, he has been the Principal Investigator of China National 863 5G Project.

Professor You served as the general chairs of IEEE WCNC 2013, IEEE VTC 2016 Spring and IEEE ICC 2019. Now he is Secretary General of the FuTURE Forum, vice Chair of China IMT-2020 (5G) Promotion Group, vice Chair of China National Mega Project on New Generation Mobile Network. He was the recipient of the National 1st Class Invention Prize in 2011, and he was selected as IEEE Fellow in same year.

\end{IEEEbiography}

\begin{IEEEbiography}[{\includegraphics[width=1.1in,height=1.3in,clip,keepaspectratio]{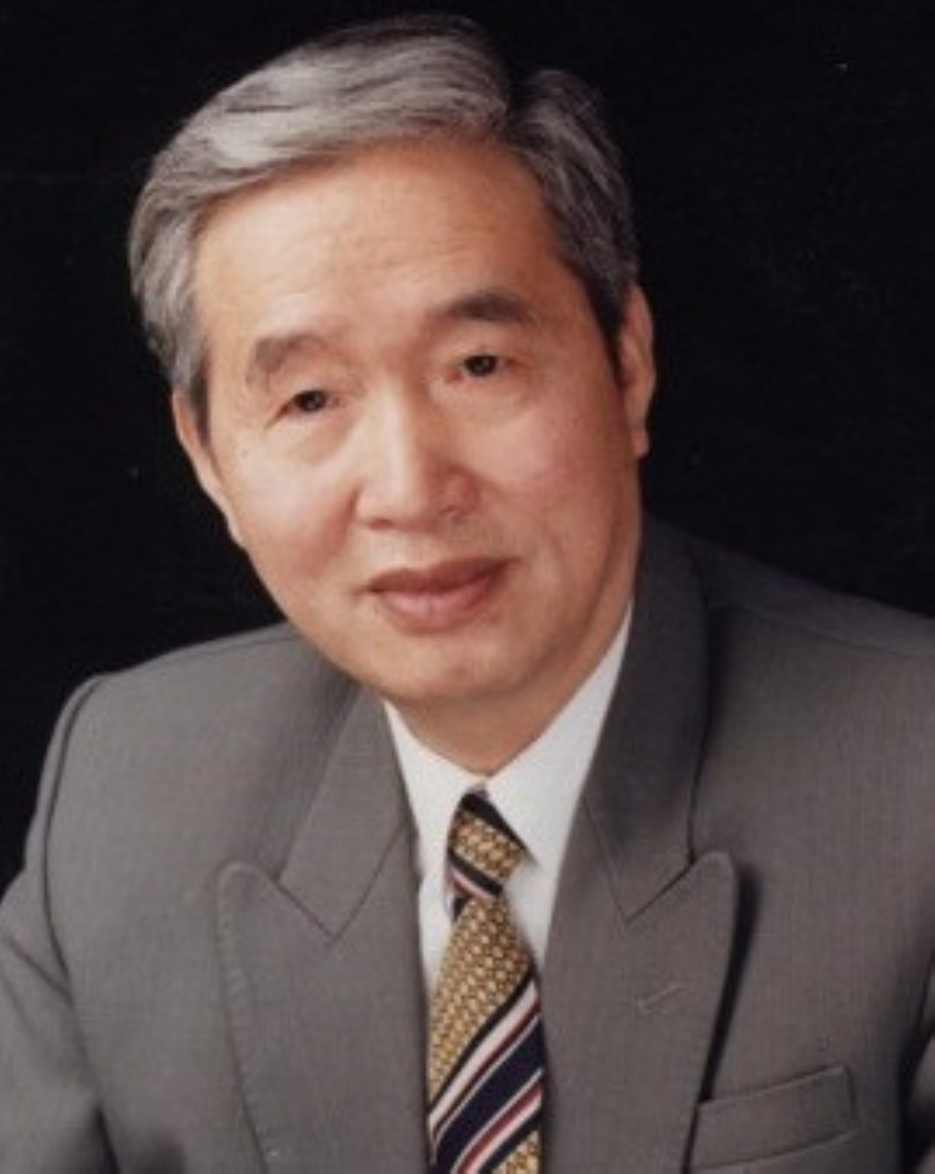}}]
{Minggao Zhang} received the BSc degree in mathematics from Wuhan University, China, in 1962.

He is currently a Distinguished Professor with the School of Information Science and Engineering, Shandong University, director of Academic Committee of China Rainbow Project Collaborative Innovation Center, director of Academic Committee of Shandong Provincial Key Lab of Wireless Communication Technologies, and a senior engineer of No. 22 Research Institute of China Electronics Technology Corporation (CETC). He has been an academician of Chinese Academy of Engineering since 1999 and is currently a Fellow of China Institute of Communications (CIC). He was a group leader of the Radio Transmission Research Group of ITU-R.

Minggao Zhang has been engaged in the research of radio propagation for decades. Many of his proposals have been adopted by international standardization organizations, including CCIR P.617-1, ITU-R P.680-3, ITU-R P.531-5, ITU-R P.529-2, and ITU-R P.676-3. In addition, he has received seven national and ministerial-level Science and Technology Progress Awards in China.
\end{IEEEbiography}
\end{document}